\documentclass[iop,apj,dvipdfmx]{emulateapj}

\slugcomment{Draft version October 13, 2016}

\shorttitle{Star Formation in ULIRGs Probed with \textit{AKARI}}
\shortauthors{Yano et al.}

\begin{document}

\title{Star Formation in Ultraluminous Infrared Galaxies
Probed with \textit{AKARI} Near-Infrared Spectroscopy}

\author{Kenichi Yano\altaffilmark{1, 2}}
\author{Takao Nakagawa\altaffilmark{2}}
\author{Naoki Isobe\altaffilmark{2, 3}}
\author{Mai Shirahata\altaffilmark{2}}
\altaffiltext{1}{Department of Physics, Graduate School of Science, The University of Tokyo,
7-3-1 Hongo, Bunkyo-ku, Tokyo 113-0033, Japan}
\altaffiltext{2}{Institute of Space and Astronautical Science, Japan Aerospace Exploration Agency,
3-1-1 Yoshinodai, Sagamihara, Kanagawa 252-5210, Japan}
\altaffiltext{3}{Department of Physics, Tokyo Institute of Technology,
2-12-1 Ookayama, Meguro-ku, Tokyo 152-8550, Japan}

\begin{abstract}
We conducted
systematic observations of the \ion{H}{1}
Br$\alpha$ line (4.05~$\mu$m)
and the polycyclic aromatic hydrocarbon
(PAH) feature (3.3~$\mu$m)
in 50 nearby ($z<0.3$) ultraluminous infrared galaxies (ULIRGs)
with \textit{AKARI}.
The Br$\alpha$ line is predicted to be the brightest
among the \ion{H}{1} lines
under high dust-extinction
conditions ($A_V>15$~mag).
The Br$\alpha$ line traces ionizing photons from OB stars
and so is used as an indicator of star formation
on the assumption of the initial mass function.
We detected the Br$\alpha$ line in 33 ULIRGs.
The luminosity of the line ($L_{\mathrm{Br}\alpha}$)
correlates well with that of
the 3.3~$\mu$m PAH emission ($L_{3.3}$).
Thus we utilize $L_{3.3}$ as an indicator of star formation
in fainter objects where the Br$\alpha$ line is undetected.
The mean $L_{\mathrm{Br}\alpha}/L_\mathrm{IR}$ ratio
in LINERs/Seyferts is
significantly lower than that in \ion{H}{2} galaxies.
This difference is reconfirmed
with the $L_{3.3}/L_\mathrm{IR}$ ratio
in the larger sample (46 galaxies).
Using the ratios,
we estimate that the contribution of starburst
in LINERs/Seyferts is $\sim67\%$,
and active galactic nuclei contribute to
the remaining $\sim33\%$.
However,
comparing the number of ionizing photons,
$Q_{\mathrm{Br}\alpha}$,
derived from $L_{\mathrm{Br}\alpha}$
with that, $Q_\mathrm{IR}$, expected from star formation rate
required to explain $L_{\mathrm{IR}}$,
we find that the mean
$Q_{\mathrm{Br}\alpha}/Q_\mathrm{IR}$ ratio
is only 55.5$\pm$7.5\%
even in \ion{H}{2} galaxies
which are thought to be
energized by pure starburst.
This deficit of ionizing photons
traced by the Br$\alpha$ line is significant
even taking heavy dust extinction into consideration.
We propose that
dust within \ion{H}{2} regions
absorbs a significant fraction of
ionizing photons.
\end{abstract}


\section{INTRODUCTION}
Ultraluminous infrared galaxies (ULIRGs)
radiate most ($\ge90\%$) of their extremely large,
quasar-like luminosities ($>10^{12}L_\odot$) as infrared dust emission
\citep{Sanders1988uig}.
The possible energy source of their enormous infrared luminosity
is starburst activities and/or
active galactic nuclei (AGN).
Morphologically, most ULIRGs exhibit merger features.
Thus, ULIRGs are thought to be
dust-enshrouded quasars formed through the
merger processes,
eventually shedding their dust to evolve into
quasars or massive ellipticals
\citep[e.g.,][]{Sanders1988uig,Sanders1996lig}.
Understanding whether
the dust-obscured energy source of ULIRGs
is dominated by starburst or AGN
is therefore important to
investigate
this merger-driven evolutionary scenario.

To identify the energy source of galaxies,
optical line ratios have been used 
to classify galaxies according to the excitation mechanism
\citep[e.g.,][]{Kim1998i1j}.
This optical classification method was promoted by
\cite{Baldwin1981cpe}, and later modified
by \cite{Veilleux1987sce}.
On the basis of two-dimensional line-intensity ratios,
such as [\ion{O}{3}]~$\lambda$5007/H$\beta$
vs [\ion{N}{2}]~$\lambda$6583/H$\alpha$,
galaxies are mainly classified into three categories;
\ion{H}{2} galaxies, Seyferts, and LINERs.
The emission lines are mainly excited by
starburst and AGN in \ion{H}{2} galaxies
and Seyferts, respectively.
The line excitation mechanisms in LINERs
are not clear and still under debate
\citep[e.g., see a review by][]{Ho2008nai}.
However, in ULIRGs,
it is difficult to distinguish starburst and AGN
through optical studies
because of their high dust extinction.
To avoid this problem,
many attempts have been made
to reveal dust-obscured energy sources
with infrared observations.

Using near-infrared observations with \textit{AKARI},
\cite{Imanishi2008si2,Imanishi2010aii}
found dust-obscured AGN signatures in a significant fraction of ULIRGs
that were optically classified as non-Seyferts.
From observations using \textit{ISO} \citep[e.g.,][]{Genzel1998wpu}
and \textit{Spitzer} \citep[e.g.,][]{Veilleux2009sqa},
the fractional contributions
of starburst and AGN
as the energy sources in ULIRGs
have been estimated with less dust extinction bias.
Their results of infrared observations,
however, rely largely on empirical relations.
For instance, many of them
use the polycyclic aromatic hydrocarbon (PAH) feature
as an indicator of starburst activities.
PAH is thought to be excited by UV photons from OB stars;
PAH emits fluorescent light at infrared wavelengths,
but its emission mechanisms are very complex
\citep{Draine2003idg}.
PAH emission has not been
theoretically related to
the number of UV photons from OB stars.
Thus, quantitative discussion about the contribution
of starburst or AGN is difficult
on the basis of these observations.

To quantitatively investigate
the energy sources of ULIRGs,
we focus on the near-infrared hydrogen recombination line
Br$\alpha$ ($n:5\rightarrow4$, $\lambda_\mathrm{rest}=4.05$ $\mu$m).
Since hydrogen is the simplest element,
the number of ionizing photons from OB stars
is calculated from
the fluxes of hydrogen recombination lines
on the basis of the photoionization
model in the case B \citep{Osterbrock2006agn}.
Star formation rates (SFRs)
can then be estimated from the number of ionizing photons
on the assumption of the initial mass function
\citep[e.g.,][]{Kennicutt2012sfi}.
Thus, we can theoretically
estimate the strength of starburst activities
with hydrogen recombination lines.
Owing to its near-infrared wavelength,
the observed flux of the Br$\alpha$ line is predicted to be the highest among
hydrogen recombination lines (i.e., H$\alpha$ or Pa$\alpha$ lines)
in the conditions of high dust extinction
(visual extinction $A_V>15$~mag)
which is expected in ULIRGs \citep[e.g.,][]{Genzel1998wpu}.
Therefore, the Br$\alpha$ line is the most suitable
for probing starburst activities in ULIRGs.

With the unique wavelength coverage
of the near-infrared 2.5--5.0~$\mu$m
spectroscopy of \textit{AKARI} \citep{Murakami2007iam,Onaka2007ici},
we succeed in systematically observing
the Br$\alpha$ line in ULIRGs,
whose wavelength is difficult to be accessed from ground-based telescopes.
The \textit{AKARI} near-infrared spectroscopy also has the unique
capability to simultaneously observe the 3.3~$\mu$m PAH emission
and the Br$\alpha$ line.
The 3.3~$\mu$m PAH emission is stronger than the Br$\alpha$ line
\citep{Imanishi2008si2,Imanishi2010aii},
and hence,
its luminosity can be
used as an indicator of star formation
for fainter objects
if we calibrate it with the
Br$\alpha$ line luminosity.

By comparing
the Br$\alpha$ line 
(or the 3.3~$\mu$m PAH emission) luminosity
with the total infrared luminosity,
we quantitatively investigate the
dust-obscured energy sources
in ULIRGs.
We assume that the Br$\alpha$ line luminosity
is proportional to the strength of the starburst
activity, i.e., SFRs.
On the other hand,
the total infrared luminosity has
contributions from the starburst
and the AGN.
Thus, the ratio of the Br$\alpha$ line luminosity
to the total infrared luminosity is expected to be
an indicator
of the contribution of starburst to
the total infrared luminosity.
In Section~2, we present our targets,
observations,
and data reduction method.
Resulting spectra and measured Br$\alpha$ line fluxes
are presented in Section~3.
We utilize the Br$\alpha$ line and
the 3.3 $\mu$m PAH emission as indicators
of star formation
and show a result of comparisons
with the total infrared luminosity in there.
Then, in Section~4, we discuss the
contribution of starburst
to the total energy from ULIRGs
by comparing the Br$\alpha$ line 
and 3.3 $\mu$m PAH emission luminosities with
the total infrared luminosity.
We summarize our study in Section~5.
Throughout this paper,
we assume that the universe is flat with
$\Omega_\mathrm{M}=0.27$,
$\Omega_\Lambda=0.73$,
and $H_0=70.4$~km~s$^{-1}$~Mpc$^{-1}$
\citep{Komatsu2011syw}.

\section{OBSERVATION AND DATA REDUCTION}
\subsection{Targets}
Our targets are selected from the \textit{AKARI} mission program
``Evolution of ultraluminous infrared galaxies
and active galactic nuclei'' (AGNUL; P.I.~T.~Nakagawa),
which aimed to investigate
the connection between ULIRGs and AGN.
The AGNUL program conducted systematic
near-infrared spectroscopic observations
of ULIRGs
in the local universe.
Among the archived data of AGNUL,
we focus on the observations
conducted during the liquid-He cool holding period
(2006 May 8--2007 August 26).
As a result, 50 near-infrared grism spectroscopic observations
of ULIRGs are selected and used for this study.
Table~\ref{log} summarizes the observation log.

\begin{deluxetable*}{lcc}
\tablecaption{Observation Log of ULIRGs.}
\tablewidth{0pt}
\tablehead{
\colhead{Object Name}&\colhead{Observation ID}&\colhead{Observation Date}
}
\startdata
IRAS 00183$-$7111&1100137.1&2007 May 2\\
IRAS 00456$-$2904&1100221.1&2007 Jun 19\\
IRAS 00482$-$2721&1100036.1&2006 Dec 21\\
IRAS 01199$-$2307&1100209.1&2007 Jul 1\\
IRAS 01298$-$0744&1100226.1&2007 Jul 10\\
IRAS 01355$-$1814&1100018.1&2007 Jan 6\\
IRAS 01494$-$1845&1100215.1&2007 Jul 10\\
IRAS 01569$-$2939&1100225.1&2007 Jul 7\\
IRAS 02480$-$3745&1100030.1&2007 Jan 14\\
IRAS 03209$-$0806&1100210.1&2007 Aug 8\\
IRAS 03521$+$0028&1100200.1&2007 Aug 19\\
IRAS 04074$-$2801&1100201.1&2007 Aug 14\\
IRAS 04313$-$1649&1100031.1&2007 Feb 22\\
IRAS 05020$-$2941&1100003.1&2007 Feb 28\\
IRAS 05189$-$2524&1100129.1&2007 Mar 8\\
IRAS 06035$-$7102&1100130.1&2007 Mar 11\\
IRAS 08572$+$3915&1100049.1&2006 Oct 29\\
IRAS 08591$+$5248&1100121.1&2007 Apr 21\\
IRAS 09320$+$6134 (UGC 5101)&1100134.1&2007 Apr 22\\
IRAS 09463$+$8141&1100004.1&2006 Oct 8\\
IRAS 09539$+$0857&1100267.1&2007 May 19\\
IRAS 10035$+$2740&1100216.1&2007 May 15\\
IRAS 10091$+$4704&1100122.1&2007 May 7\\
IRAS 10494$+$4424&1100266.1&2007 May 16\\
IRAS 10594$+$3818&1100021.1&2006 Nov 23\\
IRAS 11028$+$3130&1100006.1&2006 Nov 26\\
IRAS 11180$+$1623&1100202.1&2007 Jun 5\\
IRAS 11387$+$4116&1100269.1&2007 May 29\\
IRAS 12447$+$3721&1100022.1&2006 Dec 15\\
IRAS 12540$+$5708 (Mrk 231)&1100271.1&2007 May 30\\
IRAS 13428$+$5608 (Mrk 273)&1100273.1&2007 Jun 8\\
IRAS 13469$+$5833&1100023.1&2006 Dec 7\\
IRAS 13539$+$2920&1100235.1&2007 Jul 6\\
IRAS 14121$-$0126&1100011.1&2007 Jan 22\\
IRAS 14202$+$2615&1100212.1&2007 Jul 15\\
IRAS 14394$+$5332&1100283.1&2007 Jun 25\\
IRAS 15043$+$5754&1100213.1&2007 Jun 23\\
IRAS 16333$+$4630&1100013.1&2007 Feb 8\\
IRAS 16468$+$5200&1100249.1&2007 Aug 10\\
IRAS 16487$+$5447&1100247.1&2007 Aug 6\\
IRAS 17028$+$5817&1100248.1&2007 Aug 8\\
IRAS 17044$+$6720&1100297.1&2007 May 31\\
IRAS 17068$+$4027&1100026.1&2007 Feb 26\\
IRAS 17179$+$5444&1100253.1&2007 Aug 23\\
IRAS 19254$-$7245&1100132.1&2007 Mar 30\\
IRAS 21477$+$0502&1100207.1&2007 May 22\\
IRAS 22088$-$1831&1100214.1&2007 May 19\\
IRAS 23128$-$5919&1100294.1&2007 May 10\\
IRAS 23129$+$2548&1100015.1&2006 Dec 22\\
IRAS 23498$+$2423&1100287.1&2007 Jul 1
\enddata
\label{log}
\end{deluxetable*}

\begin{figure*}
\plottwo{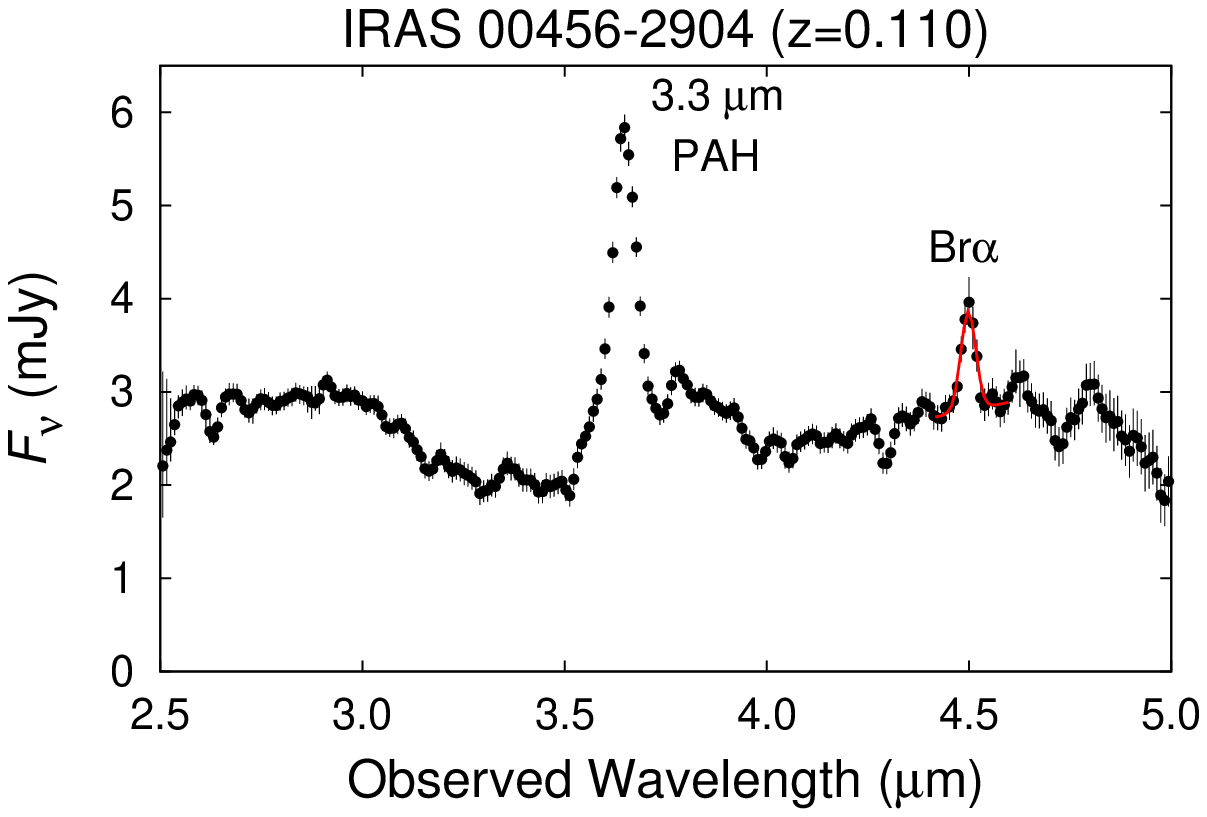}{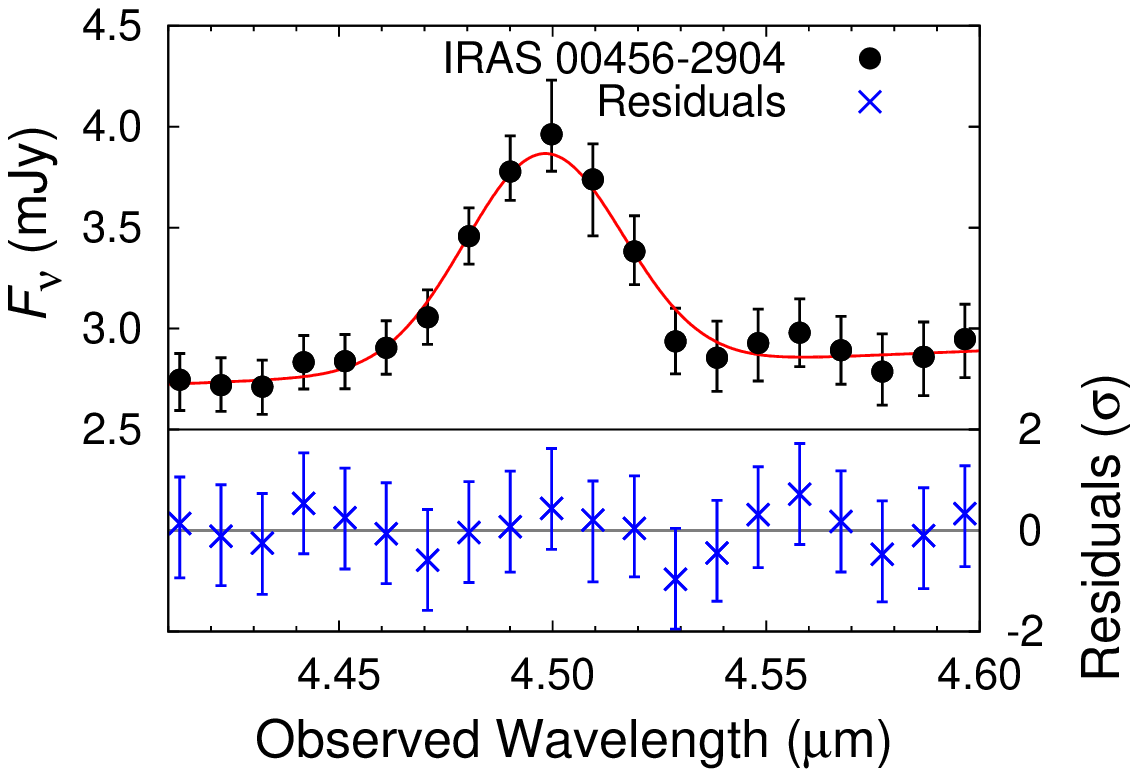}
\caption{Typical example of the
\textit{AKARI} IRC 2.5--5.0~$\mu$m spectra of ULIRGs.
The left panel presents the entire spectrum,
while the right panel shows the magnification
around the Br$\alpha$ line of the left panel.
The best-fit Gaussian profile for the Br$\alpha$ line
is plotted with the red solid line.
The residual spectrum of the best-fit is also displayed
in the left panel with the blue crosses.}
\label{spec}
\end{figure*}

The spectra of all the targets
have been reported by \cite{Imanishi2008si2,Imanishi2010aii}.
The 3.3~$\mu$m PAH feature
in these galaxies is closely investigated
in their papers,
but the Br$\alpha$ line is not discussed in detail.
Thus, we further
reduced the spectra
to systematically derive the Br$\alpha$ line fluxes
in the sample for the first time,
while the 3.3~$\mu$m PAH emission fluxes and
optical classifications
of the galaxies are taken from
\cite{Imanishi2008si2,Imanishi2010aii}.
In addition,
Table~\ref{iras} summarizes the basic information,
such as redshifts and \textit{IRAS}-based
total infrared luminosities ($L_\mathrm{IR}$)
of our target ULIRGs.

\begin{deluxetable*}{cccccccccc}
\tablecaption{Basic Information of Our Target ULIRGs.}
\tablewidth{0pt}
\tabletypesize{\tiny}
\tablehead{
\colhead{Object Name}&\colhead{$z$\tablenotemark{a}}&\colhead{$D_\mathrm{L}$\tablenotemark{b}}&\colhead{$F_{25}$\tablenotemark{c}}&\colhead{$F_{60}$\tablenotemark{c}}&\colhead{$F_{100}$\tablenotemark{c}}&\colhead{$L_\mathrm{IR}$\tablenotemark{d}}&\colhead{Optical\tablenotemark{e}}&\colhead{$F_{3.3}$\tablenotemark{f}}&\colhead{Ref.\tablenotemark{g}}\\
\colhead{}&\colhead{}&\colhead{(Mpc)}&\colhead{(Jy)}&\colhead{(Jy)}&\colhead{(Jy)}&\colhead{($10^{12}L_\odot$)}&\colhead{class}&\colhead{($10^{-14}$erg s$^{-1}$cm$^{-2}$)}&\colhead{}
}
\startdata
IRAS 00183$-$7111\tablenotemark{hi}&0.327&1717&0.13&1.20&1.19&7.56&LINER&$<1.8$&1\\
IRAS 00456$-$2904&0.110&508&0.14&2.60&3.38&1.40&\ion{H}{2}&4.0&2\\
IRAS 00482$-$2721&0.129&604&$<0.18$&1.13&1.84&1.17&LINER&1.3&3\\
IRAS 01199$-$2307&0.156&744&$<0.16$&1.61&1.37&1.22&\ion{H}{2}&0.8&2\\
IRAS 01298$-$0744\tablenotemark{j}&0.136&640&0.19&2.47&2.08&1.62&\ion{H}{2}&1.7&4\\
IRAS 01355$-$1814\tablenotemark{j}&0.192&935&0.12&1.40&1.74&2.80&\ion{H}{2}&0.6&5\\
IRAS 01494$-$1845&0.158&753&$<0.15$&1.29&1.85&1.71&Unclassified&2.2&5\\
IRAS 01569$-$2939&0.140&660&0.14&1.73&1.51&1.26&\ion{H}{2}&1.8&3\\
IRAS 02480$-$3745&0.165&790&$<0.11$&1.25&1.49&1.43&Unclassified&1.5&5\\
IRAS 03209$-$0806&0.166&798&$<0.13$&1.00&1.69&1.77&\ion{H}{2}&2.2&6\\
IRAS 03521$+$0028\tablenotemark{j}&0.152&721&0.20&2.52&3.62&3.32&LINER&2.2&7\\
IRAS 04074$-$2801\tablenotemark{j}&0.154&731&0.07&1.33&1.72&1.47&LINER&0.9&2\\
IRAS 04313$-$1649&0.268&1364&0.07&1.01&1.10&3.58&Unclassified&$<0.4$&5\\
IRAS 05020$-$2941&0.154&734&0.10&1.93&2.06&1.79&LINER&1.0&2\\
IRAS 05189$-$2524\tablenotemark{k}&0.043&187&3.47&13.25&11.84&1.55&Seyfert 2&30.0&8\\
IRAS 06035$-$7102\tablenotemark{{ilm}}\mbox{}&0.079&359&0.62&5.34&6.18&1.68&\ion{H}{2}&11.2&4\\
IRAS 08572$+$3915\tablenotemark{k}&0.058&260&1.76&7.30&4.77&1.37&LINER&$<4.2$&9\\
IRAS 08591$+$5248&0.157&750&$<0.16$&1.01&1.53&1.50&Unclassified&1.4&10\\
UGC 5101\tablenotemark{{kn}}&0.039&173&1.02&11.68&19.91&1.05&LINER&22.5&11\\
IRAS 09463$+$8141&0.156&743&$<0.07$&1.43&2.29&1.81&LINER&2.2&5\\
IRAS 09539$+$0857&0.129&603&$<0.15$&1.44&1.04&0.62&LINER&1.2&12\\
IRAS 10035$+$2740&0.166&793&$<0.17$&1.14&1.63&1.77&Unclassified&1.4&13\\
IRAS 10091$+$4704&0.246&1237&$<0.08$&1.18&1.55&3.49&LINER&0.7&5\\
IRAS 10494$+$4424&0.092&420&0.16&3.53&5.41&1.48&LINER&3.6&7\\
IRAS 10594$+$3818&0.158&753&$<0.15$&1.29&1.89&1.74&\ion{H}{2}&2.6&5\\
IRAS 11028$+$3130\tablenotemark{j}&0.199&971&0.09&1.02&1.44&2.47&LINER&0.6&12\\
IRAS 11180$+$1623&0.166&795&$<0.19$&1.19&1.60&1.80&LINER&0.5&5\\
IRAS 11387$+$4116&0.149&705&$<0.14$&1.02&1.51&1.27&\ion{H}{2}&1.9&14\\
IRAS 12447$+$3721\tablenotemark{j}&0.158&753&0.10&1.04&0.84&1.01&\ion{H}{2}&1.3&5\\
Mrk 231\tablenotemark{{ik}}&0.042&186&8.84&30.80&29.74&3.89&Seyfert 1&76.6&15\\
Mrk 273\tablenotemark{{ik}}&0.038&166&2.36&22.51&22.53&1.29&Seyfert 2&24.8&7\\
IRAS 13469$+$5833\tablenotemark{j}&0.158&752&0.04&1.27&1.73&1.43&\ion{H}{2}&1.4&1\\
IRAS 13539$+$2920&0.108&500&0.12&1.83&2.73&1.14&\ion{H}{2}&4.3&16\\
IRAS 14121$-$0126\tablenotemark{j}&0.150&712&0.11&1.39&2.07&1.84&LINER&2.9&3\\
IRAS 14202$+$2615&0.159&757&0.15&1.49&1.99&2.19&\ion{H}{2}&4.2&12\\
IRAS 14394$+$5332&0.105&481&0.35&1.95&2.39&1.40&Seyfert 2&6.2&17\\
IRAS 15043$+$5754&0.151&714&0.07&1.02&1.50&1.29&\ion{H}{2}&2.0&6\\
IRAS 16333$+$4630\tablenotemark{j}&0.191&930&0.06&1.19&2.09&2.84&LINER&2.1&7\\
IRAS 16468$+$5200&0.150&711&0.10&1.01&1.04&1.06&LINER&0.6&5\\
IRAS 16487$+$5447&0.104&477&0.20&2.88&3.07&1.23&LINER&3.3&9\\
IRAS 17028$+$5817&0.106&489&0.10&2.43&3.91&1.42&LINER&3.7&9\\
(E nucleus)&\nodata&\nodata&\nodata&\nodata&\nodata&\nodata&\ion{H}{2}&0.4&\nodata\\
(W nucleus)&\nodata&\nodata&\nodata&\nodata&\nodata&\nodata&LINER&3.3&\nodata\\
IRAS 17044$+$6720&0.135&634&0.36&1.28&0.98&1.71&LINER&2.3&18\\
IRAS 17068$+$4027&0.179&865&0.12&1.33&1.41&2.04&\ion{H}{2}&2.2&5\\
IRAS 17179$+$5444&0.147&696&0.20&1.36&1.91&2.03&Seyfert 2&1.2&5\\
IRAS 19254$-$7245\tablenotemark{{lo}}\mbox{}&0.062&275&1.34&5.30&6.70&1.54&Seyfert 2&3.8&4\\
IRAS 21477$+$0502\tablenotemark{j}&0.171&822&0.16&1.14&1.46&2.19&LINER&0.6&5\\
IRAS 22088$-$1831\tablenotemark{j}&0.170&818&0.07&1.73&1.73&1.75&\ion{H}{2}&$<0.2$&1\\
IRAS 23128$-$5919\tablenotemark{{ikm}}\mbox{}&0.045&197&1.64&10.94&10.68&1.04&\ion{H}{2}&32.0&19\\
IRAS 23129$+$2548\tablenotemark{j}&0.179&864&0.08&1.81&1.64&1.89&LINER&$<0.4$&5\\
IRAS 23498$+$2423\tablenotemark{j}&0.212&1045&0.12&1.02&1.45&3.17&Seyfert 2&$<1.9$&5
\enddata
\tablenotetext{a}{Redshift.}
\tablenotetext{b}{\mbox{}Luminosity distance calculated from $z$ using our adopted cosmology.}
\tablenotetext{c}{\textit{IRAS} fluxes at 25~$\mu$m ($F_{25}$), 60~$\mu$m ($F_{60}$),
and 100~$\mu$m ($F_{100}$).
These fluxes are taken from the \textit{IRAS} Faint Source Catalog \citep{Moshir1992ifs},
except for galaxies with additional notes.}
\tablenotetext{d}{Total infrared (3--1100~$\mu$m) luminosity calculated with
$L_\mathrm{IR}=4\pi D_\mathrm{L}^2(\xi_1\nu F_{25}+\xi_2\nu F_{60}+\xi_3\nu F_{100})$, 
$(\xi_1,\ \xi_2,\ \xi_3)=(2.403,\ -0.2454,\ 1.6381)$ \citep{Dale2002ise}.
For sources with upper limits, we follow the
method described in \cite{Imanishi2008si2,Imanishi2010aii}.
The upper and lower limits on the infrared luminosity are obtained
by assuming that the actual flux is equal to the
\textit{IRAS} upper limit and zero value, respectively,
and the average of these values is adopted as the infrared luminosity.
Since the calculation is based on our adopted cosmology,
the infrared luminosity of IRAS 09539$+$0857
is estimated to be lower than that of $10^{12}L_\odot$,
however, we treat this object as ULIRG in this paper.}
\tablenotetext{e}{Optical classification of galaxies.
These classifications are taken from \cite{Veilleux1999osi},
except for galaxies with additional notes.}
\tablenotetext{f}{Observed flux of 3.3~$\mu$m PAH emission.
These fluxes are taken from \cite{Imanishi2008si2},
except for galaxies with additional notes.}
\tablenotetext{g}
{References of redshift:
(1) \cite{Fisher1995i1j};
(2) 6dF Galaxy Survey \citep{Jones20046gs} 
Data Release (DR) 3 \citep{Jones20096gs};
(3) 2dF Galaxy Redshift Survey \citep{Colless20012gr}
Final DR \citep{Colless20032gr};
(4) \cite{Strauss1992rsi};
(5) \cite{Kim1998i1ja};
(6) Sloan Digital Sky Survey \citep[SDSS; ][]{York2000sds}
DR 1 \citep{Abazajian2003fdr};
(7) \cite{Downes1993mgm};
(8) \cite{Huchra1983sgr};
(9) \cite{Murphy2001kbs};
(10) \cite{Hwang2007uah};
(11) \cite{Rothberg2006smr};
(12) \cite{Darling2006osc};
(13) SDSS DR 6 \citep{Adelman-McCarthy2008sdr};
(14) SDSS DR 4 \citep{Adelman-McCarthy2006fdr};
(15) \cite{Carilli1998sdi};
(16) \cite{Strauss1988dig};
(17) SDSS DR 3 \citep{Abazajian2005tdr};
(18) \cite{Nagar2003acu};
(19) \cite{Lauberts1989spc}.
}
\tablenotetext{h}{Optical classification is taken from \cite{Armus1989lso}.}
\tablenotetext{i}{3.3~$\mu$m PAH emission flux is taken from \cite{Imanishi2010aii}.}
\tablenotetext{j}{\textit{IRAS} fluxes are taken from \cite{Kim1998i1ja}.}
\tablenotetext{k}{\textit{IRAS} fluxes are taken from \cite{Sanders2003irb}.}
\tablenotetext{l}{\textit{IRAS} fluxes are taken from \cite{Sanders1995ibg}.}
\tablenotetext{m}{Optical classification is taken from \cite{Duc1997sui}.}
\tablenotetext{n}{Optical classification is taken from \cite{Veilleux1995osl}.}
\tablenotetext{o}{Optical classification is taken from \cite{Mirabel1991s}.}
\label{iras}
\end{deluxetable*}

\subsection{Spectral Analysis}
The near-infrared spectroscopic
observations 
were conducted with 
the InfraRed Camera (IRC) infrared spectrograph \citep{Onaka2007ici}
onboard the \textit{AKARI}
infrared satellite \citep{Murakami2007iam}.
The 1$\times$1~arcmin$^2$ window
is used to avoid source overlap.
The pixel scale of the \textit{AKARI} IRC is
$1\farcs46\times1\farcs46$.
We used the NG grism mode \citep{Onaka2007ici} to
obtain a 2.5--5.0~$\mu$m spectrum.
The NG grism has 
a dispersion of $9.7\times10^{-3}$~$\mu$m~pix$^{-1}$
and an effective spectral resolution 
of $\lambda/\delta\lambda\sim120$
at 3.6~$\mu$m for a point source.
We employed the observing mode of IRC04,
in which one pointing comprised 8 or 9
independent frames.
Thus, though we assigned only one pointing
for each ULIRG, the effects of cosmic ray hits 
were removed.
The total net on-source exposure time is $\sim6$~min for
each ULIRG.

The data were processed using
``IRC Spectroscopy Toolkit Version 20110114,''
the standard IDL toolkit prepared for the reduction of \textit{AKARI}
IRC spectra \citep{Ohyama2007nia}.
Each frame was dark-subtracted, linearity-corrected,
and flat-field corrected.
Wavelength and flux calibrations were also made within
this toolkit.
The wavelength
calibration accuracy is taken to be
$\sim1$~pixel or $\sim10^{-2}$~$\mu$m \citep{Ohyama2007nia}.
The absolute flux calibration accuracy
is $\sim10\%$ at the central wavelength of the spectra,
and can be as large as $\sim20\%$ at the edge of the spectra \citep{Ohyama2007nia}.
We estimated a spatial extension of the object
by stacking the spectrum along the spatial direction
(i.e., perpendicular to the dispersion direction) for each source.
The measured full width at half maximum (FWHM) of the spatial profile
is typically $\sim4$--5~pixel,
which is consistent with the size of the point spread function
of \textit{AKARI} IRC in spectroscopic mode \citep{Lorente2008aip}.
We adopted an aperture width
of 5~pixel ($=7\farcs3$) along the spatial direction
for the spectrum extraction for each ULIRG.
Smoothing with a boxcar kernel of 3~pixel
in the dispersion direction
was applied to each spectrum.

\subsection{Line Fitting}

The Br$\alpha$ line at
rest-frame wavelength of
$\lambda_{\rm{rest}}=4.05$~$\mu$m
was fitted with a linear continuum and a single Gaussian profile in each spectrum.
The free parameters are the offset and the slope of the linear continuum,
the normalization of the Gaussian profile, and the central wavelength.
The line width is fixed
at the spatial width of each object 
(FWHM $\sim4$--5~pixel).
Here, we assumed that
the spectral resolution is determined by the size of each object
because the observations employ slitless spectroscopy,
and the intrinsic line widths are narrower than
the $\sim3000$~km~s$^{-1}$~$\Delta v$ resolution.
The range of wavelengths
used for the fitting was determined to
satisfactorily reproduce
the continuum emission
and is typically $\pm0.15$~$\mu$m around the central
wavelength of the Br$\alpha$ line.
The obtained central wavelengths of the Br$\alpha$ lines
exhibit small discrepancies from
those expected from redshifts.
The discrepancy is larger than the fitting error of typically $\sim10^{-3}$~$\mu$m
but within the wavelength calibration error of
$\sim10^{-2}$~$\mu$m \citep{Ohyama2007nia}.
Therefore, we shifted the wavelength of the entire spectrum 
so that the best-fit central wavelength of the Br$\alpha$ line
matches the redshift.
The flux of the Br$\alpha$ line
was then calculated by integrating the best-fit Gaussian profile.

\section{RESULTS}
\label{sec:res}
In this section, we present
a result of the measurement 
of the Br$\alpha$ line flux in ULIRGs.
We utilize
the Br$\alpha$ line luminosity
as an indicator of star formation
with investigating the effect of heavy dust extinction
on the Br$\alpha$ line.
We also present
a result of a comparison
of the Br$\alpha$ line with the total infrared luminosity
and the 3.3~$\mu$m PAH emission.

\subsection{Resultant Spectra and Br$\alpha$ Line Fluxes}
\label{subsec:brflux}
Figure~\ref{spec} presents an example of
2.5--5.0~$\mu$m spectra of ULIRGs
obtained with \textit{AKARI} IRC.
Among the sample of galaxies,
one source, IRAS~17028$+$5817,
has two nuclei in the source.
Since the eastern (E) and western (W) nuclei
of this galaxy are resolved with \textit{AKARI} IRC,
the spectra of the two nuclei are separately extracted.
Thus, in total, 51 spectra are obtained from the 50 observations (Tab.~\ref{log}).

Among the 51 spectra,
three sources (IRAS 00183$-$7111,
IRAS 04313$-$1649, and IRAS 10091$+$4704)
have a redshift higher than 0.24
and the Br$\alpha$ line was not observed
within the 2.5--5.0~$\mu$m wavelength range by \textit{AKARI}.
Moreover, the line fitting was not performed
for two other sources
(IRAS 21477$+$0502 and IRAS 23129$+$2548)
because their spectra suffer from spectral overlapping with other objects
and we were not able to determine the spatial extension of
the targets.
We therefore performed the line fitting to 46 objects
and detected the Br$\alpha$ line at 
the 99\% confidence level (2.5$\sigma$) in 33 objects.
In the remaining 13 objects,
we derived 2.5$\sigma$ upper-limit fluxes of the Br$\alpha$ line.
The measured Br$\alpha$ line fluxes are summarized in
Table~\ref{av}.

\begin{deluxetable*}{cccccc}
\tablecaption{Observed Br$\alpha$ Line Flux and Correction for Dust Extinction.}
\tablewidth{0pt}
\tabletypesize{\small}
\tablehead{
\colhead{Object Name}&\colhead{$F_{\mathrm{Br}\alpha}$\tablenotemark{a}}&\colhead{H$\alpha$/H$\beta$\tablenotemark{b}}&\colhead{$A_V$\tablenotemark{c}}&\colhead{$C_{\mathrm{Br}\alpha}$\tablenotemark{d}}&\colhead{Ref.\tablenotemark{e}}\\
\colhead{}&\colhead{($10^{-15}$erg s$^{-1}$cm$^{-2}$)}&\colhead{}&\colhead{(mag)}&\colhead{}&\colhead{}
}
\startdata
IRAS 00183$-$7111&\nodata\tablenotemark{f}&$2.9\pm1.0$&$0.06\pm0.98$&$1.002\pm0.032$&1\\
IRAS 00456$-$2904&$7.32\pm0.82$&$6.25\pm0.31$&$2.17\pm0.14$&$1.074\pm0.005$&2\\
IRAS 00482$-$2721&$3.53\pm0.81$&$9.09\pm0.45$&$3.22\pm0.14$&$1.111\pm0.005$&2\\
IRAS 01199$-$2307&$<4.5$&$6.67\pm0.33$&$2.35\pm0.14$&$1.080\pm0.005$&2\\
IRAS 01298$-$0744&$<2.2$&$6.67\pm0.33$&$2.35\pm0.14$&$1.080\pm0.005$&2\\
IRAS 01355$-$1814&$<2.6$&$6.25\pm0.31$&$2.17\pm0.14$&$1.074\pm0.005$&2\\
IRAS 01494$-$1845&$5.04\pm0.81$&\nodata&\nodata&\nodata&\nodata\\
IRAS 01569$-$2939&$5.71\pm0.83$&$16.67\pm0.83$&$4.91\pm0.14$&$1.175\pm0.005$&2\\
IRAS 02480$-$3745&$2.94\pm0.77$&\nodata&\nodata&\nodata&\nodata\\
IRAS 03209$-$0806&$5.59\pm0.87$&$11.11\pm0.56$&$3.78\pm0.14$&$1.132\pm0.005$&2\\
IRAS 03521$+$0028&$2.9\pm1.2$&$33.3\pm8.3$&$6.85\pm0.70$&$1.252\pm0.023$&2\\
IRAS 04074$-$2801&$<2.1$&$9.09\pm0.45$&$3.22\pm0.14$&$1.111\pm0.005$&2\\
IRAS 04313$-$1649&\nodata\tablenotemark{f}&$16.8\pm3.5$&$4.94\pm0.58$&$1.176\pm0.019$&3\\
IRAS 05020$-$2941&$2.08\pm0.74$&$6.25\pm0.31$&$2.17\pm0.14$&$1.074\pm0.005$&2\\
IRAS 05189$-$2524&$17.4\pm6.9$&$8.3\pm2.1$&$2.98\pm0.70$&$1.103\pm0.023$&2\\
IRAS 06035$-$7102&$7.1\pm1.7$&$13.4\pm4.4$&$4.30\pm0.93$&$1.151\pm0.030$&4\tablenotemark{g}\\
IRAS 08572$+$3915&$<14.9$&$20.8\pm1.0$&$5.52\pm0.14$&$1.199\pm0.005$&2\tablenotemark{h}\\
IRAS 08591$+$5248&$2.61\pm0.75$&$6.09\pm0.88$&$2.10\pm0.41$&$1.071\pm0.013$&5\\
UGC 5101&$<20.4$&$17.29\pm0.95$&$5.02\pm0.15$&$1.179\pm0.005$&5\\
IRAS 09463$+$8141&$3.00\pm0.90$&$25.0\pm6.3$&$6.05\pm0.70$&$1.219\pm0.023$&2\\
IRAS 09539$+$0857&$2.11\pm0.64$&$10.00\pm0.50$&$3.49\pm0.14$&$1.121\pm0.005$&2\\
IRAS 10035$+$2740&$<2.6$&$8.5\pm1.1$&$3.03\pm0.35$&$1.105\pm0.012$&5\\
IRAS 10091$+$4704&\nodata\tablenotemark{f}&$12.50\pm0.63$&$4.11\pm0.14$&$1.144\pm0.005$&2\\
IRAS 10494$+$4424&$11.59\pm0.68$&$9.09\pm0.45$&$3.22\pm0.14$&$1.111\pm0.005$&2\\
IRAS 10594$+$3818&$6.7\pm1.1$&$5.88\pm0.29$&$2.00\pm0.14$&$1.068\pm0.005$&2\\
IRAS 11028$+$3130&$<6.7$&$3.45\pm0.17$&$0.51\pm0.14$&$1.017\pm0.005$&2\\
IRAS 11180$+$1623&$<4.1$&$10.00\pm0.50$&$3.49\pm0.14$&$1.121\pm0.005$&2\\
IRAS 11387$+$4116&$5.16\pm0.71$&$12.50\pm0.63$&$4.11\pm0.14$&$1.144\pm0.005$&2\\
IRAS 12447$+$3721&$4.73\pm0.85$&$5.26\pm0.26$&$1.69\pm0.14$&$1.057\pm0.005$&2\\
Mrk 231&$<99.7$&$5.2\pm2.2$&$1.6\pm1.2$&$1.055\pm0.039$&6\\
Mrk 273&$38.8\pm2.3$&$10.00\pm0.50$&$3.49\pm0.14$&$1.121\pm0.005$&2\\
IRAS 13469$+$5833&$5.0\pm1.2$&$11.11\pm0.56$&$3.78\pm0.14$&$1.132\pm0.005$&2\\
IRAS 13539$+$2920&$12.5\pm1.6$&$14.29\pm0.71$&$4.48\pm0.14$&$1.158\pm0.005$&2\\
IRAS 14121$-$0126&$3.99\pm0.87$&$12.50\pm0.63$&$4.11\pm0.14$&$1.144\pm0.005$&2\\
IRAS 14202$+$2615&$7.69\pm0.90$&$6.25\pm0.31$&$2.17\pm0.14$&$1.074\pm0.005$&2\\
IRAS 14394$+$5332&$7.93\pm0.76$&$7.69\pm0.38$&$2.75\pm0.14$&$1.095\pm0.005$&2\\
IRAS 15043$+$5754&$1.95\pm0.50$&$7.69\pm0.38$&$2.75\pm0.14$&$1.095\pm0.005$&2\\
IRAS 16333$+$4630&$4.16\pm0.92$&$8.33\pm0.42$&$2.98\pm0.14$&$1.103\pm0.005$&2\\
IRAS 16468$+$5200&$<3.9$&$9.20\pm0.35$&$3.25\pm0.11$&$1.113\pm0.003$&2\tablenotemark{h}\\
IRAS 16487$+$5447&$9.33\pm0.72$&$5.00\pm0.25$&$1.55\pm0.14$&$1.052\pm0.005$&2\\
IRAS 17028$+$5817E&$2.38\pm0.74$&$5.56\pm0.28$&$1.84\pm0.14$&$1.062\pm0.005$&2\\
IRAS 17028$+$5817W&$5.15\pm0.77$&$14.29\pm0.71$&$4.48\pm0.14$&$1.158\pm0.005$&2\\
IRAS 17044$+$6720&$6.7\pm1.7$&$7.14\pm0.36$&$2.55\pm0.14$&$1.087\pm0.005$&2\\
IRAS 17068$+$4027&$8.79\pm0.95$&$12.50\pm0.63$&$4.11\pm0.14$&$1.144\pm0.005$&2\\
IRAS 17179$+$5444&$4.0\pm1.1$&$10.00\pm0.50$&$3.49\pm0.14$&$1.121\pm0.005$&2\\
IRAS 19254$-$7245&$15.7\pm2.5$&$9.8\pm1.6$&$3.43\pm0.45$&$1.119\pm0.015$&1\\
IRAS 21477$+$0502&\nodata\tablenotemark{i}&$5.88\pm0.29$&$2.00\pm0.14$&$1.068\pm0.005$&2\\
IRAS 22088$-$1831&$<2.2$&$4.76\pm0.24$&$1.41\pm0.14$&$1.047\pm0.005$&2\\
IRAS 23128$-$5919&$64.2\pm1.5$&$7.95\pm0.80$&$2.85\pm0.28$&$1.098\pm0.009$&4\tablenotemark{h}\\
IRAS 23129$+$2548&\nodata\tablenotemark{i}&$12.50\pm0.63$&$4.11\pm0.14$&$1.144\pm0.005$&2\\
IRAS 23498$+$2423&$<3.3$&$7.14\pm0.36$&$2.55\pm0.14$&$1.087\pm0.005$&2
\enddata
\tablenotetext{a}{Observed Br$\alpha$ line flux obtained with our \textit{AKARI} result.}
\tablenotetext{b}{Flux ratio of H$\alpha$ and H$\beta$ lines taken from literatures.}
\tablenotetext{c}{Visual extinction derived from H$\alpha$/H$\beta$ line ratio.}
\tablenotetext{d}{Dust-extinction correction factor for Br$\alpha$ line flux.}
\tablenotetext{e}
{References of optical line fluxes:
(1) \cite{Buchanan2006rei};
(2) \cite{Veilleux1999osi};
(3) \cite{Rupke2008oal};
(4) \cite{Duc1997sui};
(5) \cite{Hou2009uli};
(6) \cite{Lipari1994gwe}.}
\tablenotetext{f}{Redshift is larger than 0.24 and Br$\alpha$ line is not observed within 2.5--5.0~$\mu$m wavelength range.}
\tablenotetext{g}{Value of W nucleus.}
\tablenotetext{h}{Sum of two nuclei.}
\tablenotetext{i}{Spectrum suffered from source overlapping.}
\label{av}
\end{deluxetable*}

\subsection{Effect of Dust Extinction on Br$\alpha$ Line}
\label{subsec:hde}

In the standard extinction curve
\citep[e.g.,][]{Draine2003idg},
dust extinction is lower
at longer wavelengths.
Thus the Br$\alpha$ line
is less affected by
dust extinction owing to its infrared wavelength (4.05 $\mu$m)
than the optical Balmer lines and
even the infrared Pa$\alpha$ and
Br$\gamma$ lines ($\sim$2~$\mu$m) widely observed
from the ground.
Furthermore,
from observations of the Galactic center,
\cite{Fritz2011ldi} recently showed that
extinction at 4~$\mu$m 
was about the same as that at $\sim$7--8~$\mu$m
($A_{7.5\,\mu\mathrm{m}}/A_{4\,\mu\mathrm{m}}\sim0.8$,
where $A_\lambda$ is extinction in magnitude at a wavelength of $\lambda$)
and was even lower than that at $\sim$9--20~$\mu$m
($A_{12.4\,\mu\mathrm{m}}/A_{4\,\mu\mathrm{m}}\sim1.3$).
This means that
the effect of dust extinction
on the Br$\alpha$ line is
compatible with (or even lower than)
that on other
star formation indicators
in the mid-infrared wavelength
such as the 7.7~$\mu$m PAH feature
or the [\ion{Ne}{2}] 12.8~$\mu$m line
widely observed with \textit{Spitzer}
\citep[e.g.,][]{Veilleux2009sqa}.
Thus we conclude that the Br$\alpha$ line is
one of the best tracers of star formation
in the near- to mid-infrared wavelength range.
However, especially in ULIRGs, which harbor vast amounts of dust,
dust extinction could pose a problem
even if we use the Br$\alpha$ line.

As an indicator of extinction,
the optical H$\alpha$/H$\beta$ line ratio
is widely used
\citep[e.g.,][]{Kim1998i1j}.
The H$\alpha$/H$\beta$ line ratio
taken from literature
and the inferred visual extinction in our sample
are summarized in Table~\ref{av}.
We assume an intrinsic line ratio for H$\alpha$/H$\beta$
of 2.87 \citep[][case B with $T=10000$~K and low-density limit]{Osterbrock2006agn}.
For the extinction law, the Milky way dust model of \cite{Draine2003idg}
is used ($A_{\mathrm{H}\alpha}=0.776A_V$,
$A_{\mathrm{H}\beta}=1.17A_V$,
and $A_{\mathrm{Br}\alpha}=3.56\times10^{-2}A_V$).
The extinction corrected
luminosity of the Br$\alpha$ line
($L_{\mathrm{Br}\alpha}$) is
summarized in Table~\ref{lumi}.

\begin{deluxetable}{ccc}
\tablecaption{Luminosity of Br$\alpha$ Lines.}
\tablewidth{0pt}
\tablehead{
\colhead{Object Name}&\colhead{$L_{\mathrm{Br}\alpha}$\tablenotemark{a}}&\colhead{$L_{\mathrm{Br}\alpha}/L_{\mathrm{IR}}$}\\
\colhead{}&\colhead{($10^7L_\odot$)}&\colhead{($10^{-5}$)}
}
\startdata
IRAS 00183$-$7111&\nodata&\nodata\\
IRAS 00456$-$2904&$6.32\pm0.71$&$4.50\pm0.51$\\
IRAS 00482$-$2721&$4.5\pm1.0$&$3.83\pm0.88$\\
IRAS 01199$-$2307&$<8.4$&$<6.9$\\
IRAS 01298$-$0744&$<3.1$&$<1.9$\\
IRAS 01355$-$1814&$<7.7$&$<2.7$\\
IRAS 01494$-$1845&$8.9\pm1.4$&$5.21\pm0.84$\\
IRAS 01569$-$2939&$9.1\pm1.3$&$7.2\pm1.1$\\
IRAS 02480$-$3745&$5.7\pm1.5$&$4.0\pm1.0$\\
IRAS 03209$-$0806&$12.6\pm2.0$&$7.1\pm1.1$\\
IRAS 03521$+$0028&$5.9\pm2.4$&$1.78\pm0.71$\\
IRAS 04074$-$2801&$<3.9$&$<2.6$\\
IRAS 04313$-$1649&\nodata&\nodata\\
IRAS 05020$-$2941&$3.8\pm1.3$&$2.10\pm0.75$\\
IRAS 05189$-$2524&$2.10\pm0.84$&$1.35\pm0.54$\\
IRAS 06035$-$7102&$3.28\pm0.78$&$1.96\pm0.47$\\
IRAS 08572$+$3915&$<3.7$&$<2.7$\\
IRAS 08591$+$5248&$4.9\pm1.4$&$3.26\pm0.94$\\
UGC 5101&$<2.2$&$<2.1$\\
IRAS 09463$+$8141&$6.3\pm1.9$&$3.5\pm1.0$\\
IRAS 09539$+$0857&$2.68\pm0.81$&$4.3\pm1.3$\\
IRAS 10035$+$2740&$<5.7$&$<3.2$\\
IRAS 10091$+$4704&\nodata&\nodata\\
IRAS 10494$+$4424&$7.08\pm0.42$&$4.80\pm0.28$\\
IRAS 10594$+$3818&$12.6\pm2.1$&$7.2\pm1.2$\\
IRAS 11028$+$3130&$<20.0$&$<8.1$\\
IRAS 11180$+$1623&$<9.0$&$<5.0$\\
IRAS 11387$+$4116&$9.1\pm1.3$&$7.22\pm0.99$\\
IRAS 12447$+$3721&$8.8\pm1.6$&$8.7\pm1.6$\\
Mrk 231&$<11.3$&$<2.9$\\
Mrk 273&$3.72\pm0.22$&$2.88\pm0.17$\\
IRAS 13469$+$5833&$10.0\pm2.3$&$7.0\pm1.6$\\
IRAS 13539$+$2920&$11.3\pm1.4$&$9.9\pm1.2$\\
IRAS 14121$-$0126&$7.2\pm1.6$&$3.93\pm0.86$\\
IRAS 14202$+$2615&$14.8\pm1.7$&$6.73\pm0.79$\\
IRAS 14394$+$5332&$6.26\pm0.60$&$4.47\pm0.43$\\
IRAS 15043$+$5754&$3.40\pm0.87$&$2.63\pm0.67$\\
IRAS 16333$+$4630&$12.4\pm2.7$&$4.35\pm0.97$\\
IRAS 16468$+$5200&$<6.8$&$<6.4$\\
IRAS 16487$+$5447&$6.97\pm0.54$&$5.67\pm0.44$\\
IRAS 17028$+$5817&$6.32\pm0.88$&$4.45\pm0.62$\\
(E nucleus)&$1.88\pm0.59$&\nodata\\
(W nucleus)&$4.44\pm0.66$&\nodata\\
IRAS 17044$+$6720&$9.1\pm2.3$&$5.3\pm1.4$\\
IRAS 17068$+$4027&$23.4\pm2.5$&$11.5\pm1.2$\\
IRAS 17179$+$5444&$6.8\pm1.9$&$3.34\pm0.92$\\
IRAS 19254$-$7245&$4.16\pm0.66$&$2.71\pm0.43$\\
IRAS 21477$+$0502&\nodata&\nodata\\
IRAS 22088$-$1831&$<4.8$&$<2.7$\\
IRAS 23128$-$5919&$8.49\pm0.22$&$8.17\pm0.21$\\
IRAS 23129$+$2548&\nodata&\nodata\\
IRAS 23498$+$2423&$<12.1$&$<3.8$
\enddata
\tablenotetext{a}{{}Luminosity of the Br$\alpha$ line ($L_{\mathrm{Br}\alpha}$).
Dust extinction is corrected with the H$\alpha$/H$\beta$ line ratio.}
\label{lumi}
\end{deluxetable}

The visual extinction
derived from the optical H$\alpha$/H$\beta$ line ratio
is typically a few mag,
and the correction factor for the flux of the Br$\alpha$ 
line, $C_{\mathrm{Br}\alpha}$,
is at most $\sim1.3$ (Table~\ref{av}).
If dust obscuration of starburst in these ULIRGs is so heavy, however,
optical observations would trace only the outer region of the obscured starburst,
and the visual extinction derived from the H$\alpha$/H$\beta$ line ratio
would be regarded as a kind of lower limit.
For instance,
\cite{Genzel1998wpu} reported visual extinction
of 5--50~mag in ULIRGs based on the \textit{ISO}
mid-infrared spectroscopic observations.
Here we discuss the uncertainty arising from such
heavy dust obscuration.

With our \textit{AKARI} result,
we are able to estimate visual extinction
using the Br$\beta$ line
($n:6\rightarrow4$, $\lambda_\mathrm{rest}=2.63$~$\mu$m)
observed simultaneously with the Br$\alpha$ line,
although it is difficult to measure
the flux of the Br$\beta$ line
because of its faintness.
In one source among our sample, IRAS~00456$-$2904 (Figure~\ref{spec}),
we marginally detect the Br$\beta$ line
and measure its flux to be
$(2.3\pm1.0)\times10^{-15}$~erg~cm$^{-2}$~s$^{-2}$.
This yields the Br$\beta$/Br$\alpha$ line ratio
of $0.31\pm0.15$.
Adopting an intrinsic line ratio for Br$\beta$/Br$\alpha$
of 0.57 \citep{Osterbrock2006agn}
and $A_{\mathrm{Br}\beta}=8.09\times10^{-2}A_V$
\citep{Draine2003idg},
we derive visual extinction of $14\pm11$~mag
from the Br$\beta$/Br$\alpha$ line ratio.
This corresponds to a correction factor for the flux
of the Br$\alpha$ line of $1.6\pm0.5$.
Thus, in this source, extinction
heavier than that derived from
the H$\alpha$/H$\beta$ line ratio
($A_V=2.17\pm0.14$~mag)
is indicated.

To investigate the effect of 
heavy dust extinction
with the entire sample,
we focus on the optical depth of
the 9.7~$\mu$m silicate absorption feature ($\tau_{9.7}$)
derived from the \textit{Spitzer} observations.
Owing to the mid-infrared wavelength,
the 9.7~$\mu$m silicate absorption feature
can probe dust extinction in
heavily obscured regions.
Table~\ref{tab:sfcav}
summarizes $\tau_{9.7}$ taken from the literature
(references are also listed therein).
We convert $\tau_{9.7}$ to visual extinction
$A_V^{9.7}$ using the following relation;
$A_V^{9.7}\ (\mathrm{mag})=(1.08/0.087)\tau_{9.7}$
\citep{Imanishi2007sil}.
We regard $A_V^{9.7}$
as the most extreme extinction in the sample.
The mean value of $A_V^{9.7}$ is
$\sim23$ mag in our sample.
This corresponds to the correction factor
for the flux of the Br$\alpha$ line of $\sim2.1$.
Thus the effect of heavy dust
extinction to the Br$\alpha$ line flux
is about a factor of two even in the extreme cases.

\begin{deluxetable*}{cccccc}
\tablecaption{Luminosity of 3.3~$\mu$m PAH Emission and 9.7~$\mu$m Silicate Absorption.}
\tablewidth{0pt}
\tablehead{
\colhead{Object Name}&\colhead{$\tau_{9.7}$\tablenotemark{a}}&\colhead{$A_V^{9.7}$\tablenotemark{b}}&\colhead{$L_{3.3}$\tablenotemark{c}}&\colhead{$L_{3.3}/L_{\mathrm{IR}}$}&\colhead{Ref.\tablenotemark{d}}\\
\colhead{}&\colhead{}&\colhead{(mag)}&\colhead{($10^8L_\odot$)}&\colhead{($10^{-4}$)}&\colhead{}
}
\startdata
IRAS 00183$-$7111&$3.1\pm0.16$&$38.5\pm1.9$&$<16.6$&$<2.19$&1\\
IRAS 00456$-$2904&$1.2\pm0.12$&$14.9\pm1.5$&$3.6\pm1.1$&$2.55\pm0.77$&2\\
IRAS 00482$-$2721&$2.1\pm0.11$&$26.1\pm1.3$&$1.73\pm0.52$&$1.49\pm0.45$&2\\
IRAS 01199$-$2307&$2.4\pm0.12$&$29.8\pm1.5$&$1.55\pm0.46$&$1.27\pm0.38$&3\\
IRAS 01298$-$0744&$4.0\pm0.20$&$49.7\pm2.5$&$2.44\pm0.73$&$1.51\pm0.45$&2\\
IRAS 01355$-$1814&$2.4\pm0.12$&$29.8\pm1.5$&$1.82\pm0.55$&$0.65\pm0.19$&3\\
IRAS 01494$-$1845&$1.6\pm0.16$&$19.9\pm2.0$&$3.9\pm1.2$&$2.28\pm0.68$&3\\
IRAS 01569$-$2939&$2.8\pm0.14$&$34.8\pm1.7$&$3.11\pm0.93$&$2.46\pm0.74$&2\\
IRAS 02480$-$3745&$1.4\pm0.14$&$17.4\pm1.7$&$2.92\pm0.88$&$2.04\pm0.61$&4\\
IRAS 03209$-$0806&$1.0\pm0.10$&$12.4\pm1.2$&$5.3\pm1.6$&$2.96\pm0.89$&4\\
IRAS 03521$+$0028&$1.3\pm0.13$&$16.1\pm1.6$&$5.0\pm1.5$&$1.50\pm0.45$&3\\
IRAS 04074$-$2801&$3.0\pm0.15$&$37.3\pm1.9$&$1.75\pm0.53$&$1.19\pm0.36$&4\\
IRAS 04313$-$1649&$2.8\pm0.14$&$34.8\pm1.7$&$<3.0$&$<0.82$&3\\
IRAS 05020$-$2941&$2.4\pm0.12$&$29.8\pm1.5$&$1.87\pm0.56$&$1.05\pm0.31$&4\\
IRAS 05189$-$2524&$0.3\pm0.02$&$3.92\pm0.21$&$3.8\pm1.1$&$2.45\pm0.74$&5\\
IRAS 06035$-$7102&$2.9\pm0.15$&$36.0\pm1.8$&$5.6\pm1.7$&$3.3\pm1.0$&1\\
IRAS 08572$+$3915&$3.8\pm0.19$&$47.2\pm2.4$&$<1.2$&$<0.84$&2\\
IRAS 08591$+$5248&$1.0\pm0.10$&$12.4\pm1.2$&$2.72\pm0.82$&$1.81\pm0.54$&4\\
UGC 5101&$1.6\pm0.08$&$20.1\pm1.0$&$2.68\pm0.80$&$2.55\pm0.77$&5\\
IRAS 09463$+$8141&$2.0\pm0.20$&$24.8\pm2.5$&$5.1\pm1.5$&$2.82\pm0.85$&2\\
IRAS 09539$+$0857&$3.5\pm0.18$&$43.5\pm2.1$&$1.61\pm0.48$&$2.59\pm0.78$&2\\
IRAS 10035$+$2740&$2.0\pm0.20$&$24.8\pm2.5$&$3.18\pm0.96$&$1.79\pm0.54$&3\\
IRAS 10091$+$4704&$2.5\pm0.13$&$31.0\pm1.6$&$4.1\pm1.2$&$1.17\pm0.35$&3\\
IRAS 10494$+$4424&$1.7\pm0.17$&$21.1\pm2.1$&$2.32\pm0.70$&$1.57\pm0.47$&2\\
IRAS 10594$+$3818&$1.0\pm0.10$&$12.4\pm1.2$&$5.1\pm1.5$&$2.91\pm0.87$&4\\
IRAS 11028$+$3130&$2.5\pm0.13$&$31.0\pm1.6$&$1.81\pm0.54$&$0.73\pm0.22$&3\\
IRAS 11180$+$1623&$2.0\pm0.20$&$24.8\pm2.5$&$1.17\pm0.35$&$0.65\pm0.19$&3\\
IRAS 11387$+$4116&$1.1\pm0.11$&$13.7\pm1.4$&$3.6\pm1.1$&$2.84\pm0.85$&2\\
IRAS 12447$+$3721&$1.7\pm0.17$&$21.1\pm2.1$&$2.50\pm0.75$&$2.46\pm0.74$&4\\
Mrk 231&$0.6\pm0.03$&$7.98\pm0.43$&$8.9\pm2.7$&$2.29\pm0.70$&5\\
Mrk 273&$1.7\pm0.09$&$21.7\pm1.1$&$2.52\pm0.76$&$1.95\pm0.58$&5\\
IRAS 13469$+$5833&$1.7\pm0.17$&$21.1\pm2.1$&$2.97\pm0.89$&$2.08\pm0.62$&3\\
IRAS 13539$+$2920&$1.6\pm0.16$&$19.9\pm2.0$&$4.2\pm1.3$&$3.7\pm1.1$&2\\
IRAS 14121$-$0126&$1.3\pm0.13$&$16.1\pm1.6$&$5.6\pm1.7$&$3.05\pm0.92$&4\\
IRAS 14202$+$2615&$0.7\pm0.07$&$8.69\pm0.86$&$8.3\pm2.5$&$3.8\pm1.1$&4\\
IRAS 14394$+$5332&\nodata&\nodata&$5.1\pm1.5$&$3.7\pm1.1$&\nodata\\
IRAS 15043$+$5754&$1.4\pm0.14$&$17.4\pm1.7$&$3.6\pm1.1$&$2.81\pm0.84$&4\\
IRAS 16333$+$4630&$1.3\pm0.13$&$16.1\pm1.6$&$6.5\pm2.0$&$2.31\pm0.69$&3\\
IRAS 16468$+$5200&$2.5\pm0.13$&$31.0\pm1.6$&$1.11\pm0.33$&$1.04\pm0.31$&2\\
IRAS 16487$+$5447&$1.8\pm0.18$&$22.4\pm2.2$&$2.53\pm0.76$&$2.06\pm0.62$&2\\
IRAS 17028$+$5817&$1.5\pm0.15$&$18.6\pm1.9$&$3.38\pm0.92$&$2.38\pm0.65$&2\\
(E Nucleus)&\nodata&\nodata&$0.33\pm0.10$&\nodata&\nodata\\
(W Nucleus)&\nodata&\nodata&$3.06\pm0.92$&\nodata&\nodata\\
IRAS 17044$+$6720&$1.8\pm0.18$&$22.4\pm2.2$&$3.26\pm0.98$&$1.91\pm0.57$&2\\
IRAS 17068$+$4027&$1.8\pm0.18$&$22.4\pm2.2$&$6.3\pm1.9$&$3.07\pm0.92$&3\\
IRAS 17179$+$5444&\nodata&\nodata&$2.15\pm0.64$&$1.06\pm0.32$&\nodata\\
IRAS 19254$-$7245&$1.3\pm0.07$&$16.7\pm0.9$&$1.06\pm0.32$&$0.69\pm0.21$&5\\
IRAS 21477$+$0502&$0.8\pm0.08$&$9.9\pm1.0$&$1.39\pm0.42$&$0.64\pm0.19$&4\\
IRAS 22088$-$1831&$2.6\pm0.13$&$32.3\pm1.6$&$<0.45$&$<0.26$&4\\
IRAS 23128$-$5919&\nodata&\nodata&$4.4\pm1.3$&$4.3\pm1.3$&\nodata\\
IRAS 23129$+$2548&$2.6\pm0.13$&$32.3\pm1.6$&$<1.1$&$<0.60$&3\\
IRAS 23498$+$2423&$0.6\pm0.03$&$7.48\pm0.36$&$<7.3$&$<2.3$&6
\enddata
\tablenotetext{a}{Optical depth of the 9.7~$\mu$m silicate absorption.}
\tablenotetext{b}{Visual extinction derived from $\tau_{9.7}$.}
\tablenotetext{c}{{}Luminosity of the 3.3~$\mu$m PAH emission.
Dust extinction is corrected using the H$\alpha$/H$\beta$ ratio.}
\tablenotetext{d}
{References of $\tau_{9.7}$:
(1) \cite{Dartois2007cdg};
(2) \cite{Imanishi2007sil};
(3) \cite{Imanishi2009lba};
(4) \cite{Imanishi2010sil};
(5) \cite{Wu2009s53};
(6) \cite{Willett2011mip}.
}
\label{tab:sfcav}
\end{deluxetable*}

In summary,
taking the effect of heavy dust extinction into account,
we conclude that the intrinsic Br$\alpha$ line flux
is determined within an uncertainty of a factor of two underestimation.
Allowing for this uncertainty, 
we adopt and apply the dust extinction correction estimated
from the H$\alpha$/H$\beta$ line ratio
to the Br$\alpha$ line flux
as the minimum correction
because the H$\alpha$/H$\beta$ line ratio
is available for all our targets.
At least half of ionizing photons
originate from dust obscured regions are expected to be traced
by the Br$\alpha$ line,
and thus we utilize the Br$\alpha$ line luminosity
as a good indicator of star formation.

\subsection{Comparison to Other Star-Formation Indicators}
\label{subsec:comp}

The total infrared luminosity is
widely used as an indicator of star formation
in star-forming galaxies where AGN do not contaminate
the luminosity \citep[e.g.,][]{Kennicutt2012sfi}.
Thus we expect that $L_{\mathrm{Br}\alpha}$
is correlated with $L_\mathrm{IR}$
if $L_\mathrm{IR}$ is solely governed
by star formation.
Here we compare $L_{\mathrm{Br}\alpha}$
with $L_\mathrm{IR}$.

\begin{figure*}
\plottwo{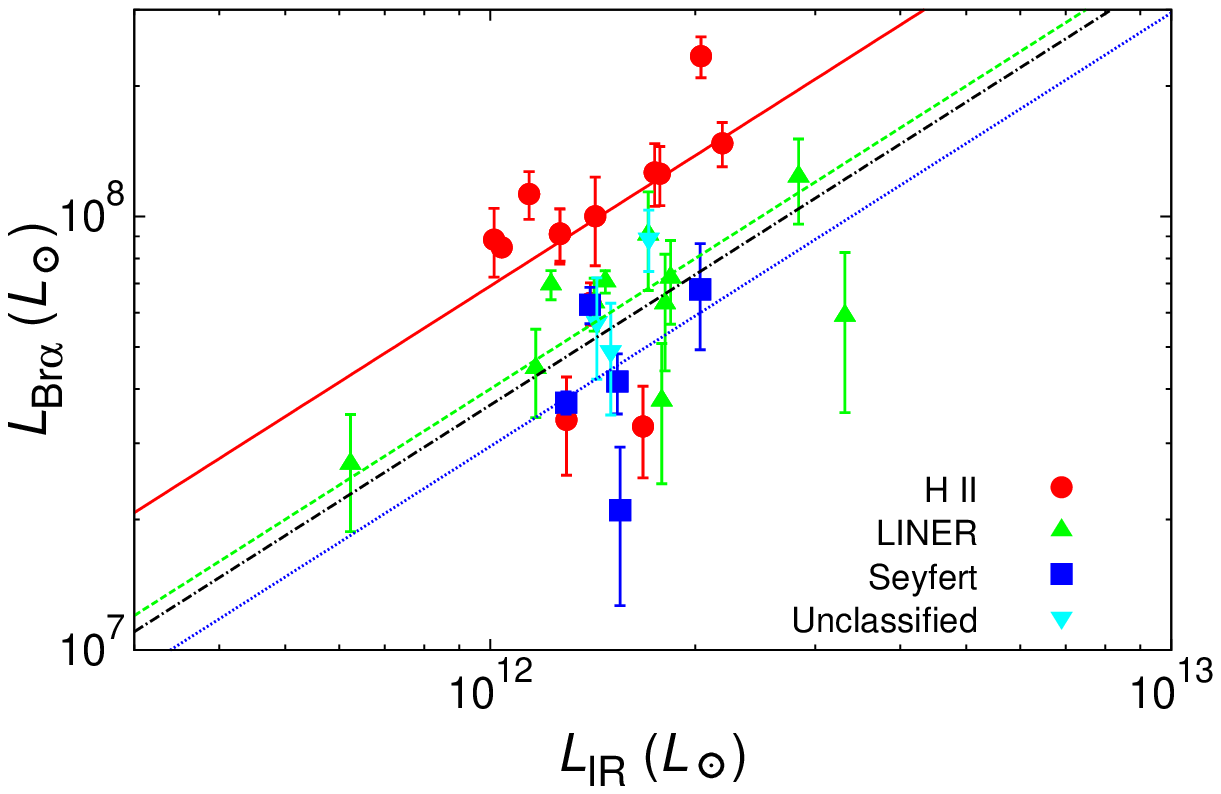}{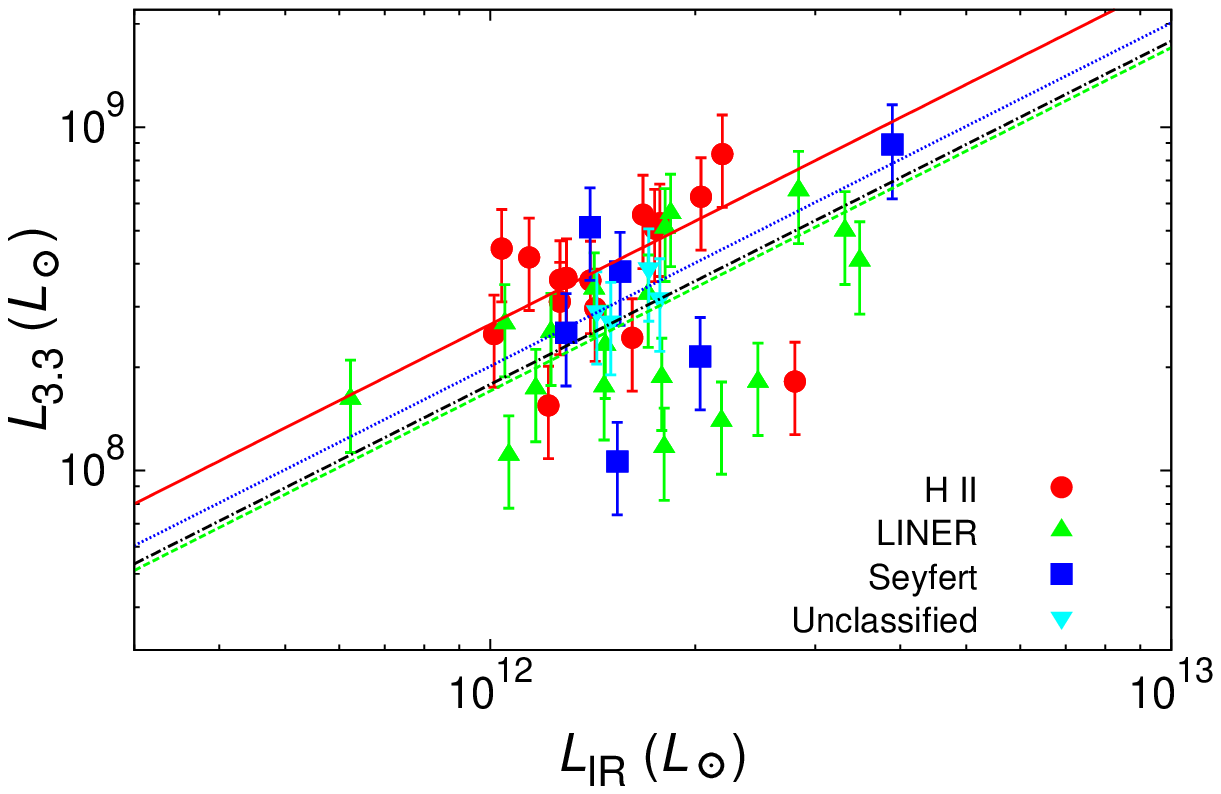}
\caption{Comparison of the Br$\alpha$ line luminosity
($L_{\mathrm{Br}\alpha}$; left)
and the 3.3 $\mu$m PAH emission luminosity ($L_{3.3}$; right)
with the total infrared luminosity ($L_{\mathrm{IR}}$).
The symbols and colors represent the optical classifications
of the galaxy.
The red solid, green dashed, blue dotted,
and black dashed--dotted lines indicate
the mean $L_{\mathrm{Br}\alpha}/L_\mathrm{IR}$ 
($L_{3.3}/L_\mathrm{IR}$) ratio
for \ion{H}{2} galaxies,
LINERs,
Seyferts,
and the combination of LINERs and Seyferts,
respectively.
}
\label{fig:brvsir}
\end{figure*}

The left panel of
Figure~\ref{fig:brvsir} shows the comparison
of $L_{\mathrm{Br}\alpha}$
with $L_{\mathrm{IR}}$ for the 33 objects
in which the Br$\alpha$ line is detected.
We find that the galaxies do not follow a single relation
and show a scatter in this plot.
The correlation coefficient between $L_\mathrm{IR}$ and $L_{\mathrm{Br}\alpha}$
is calculated to be 0.116,
which yields the probability to obtain such a correlation coefficient
by chance of 52.6\%.
This indicates no significant correlation between them.

Within the spectral window
of our \textit{AKARI} observation,
the 3.3~$\mu$m PAH emission,
which traces UV photons from OB stars
and is also expected to be used as an indicator of
star formation,
is observed simultaneously with the Br$\alpha$ line.
Although complex emission mechanisms make
it difficult to quantitatively connect 
the 3.3~$\mu$m PAH emission luminosity ($L_{3.3}$)
with the number of ionizing photons,
we also expect a correlation between
$L_{3.3}$ and $L_{\mathrm{Br}\alpha}$.

\begin{figure*}
\plottwo{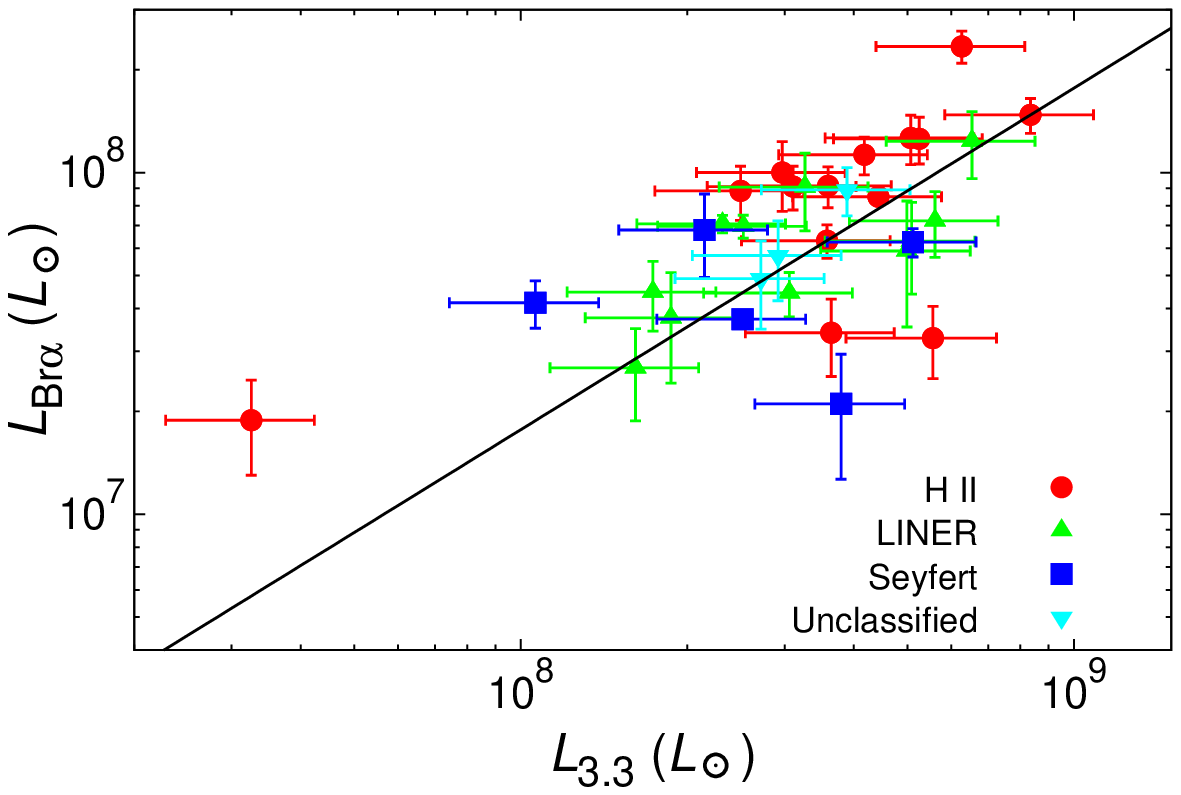}{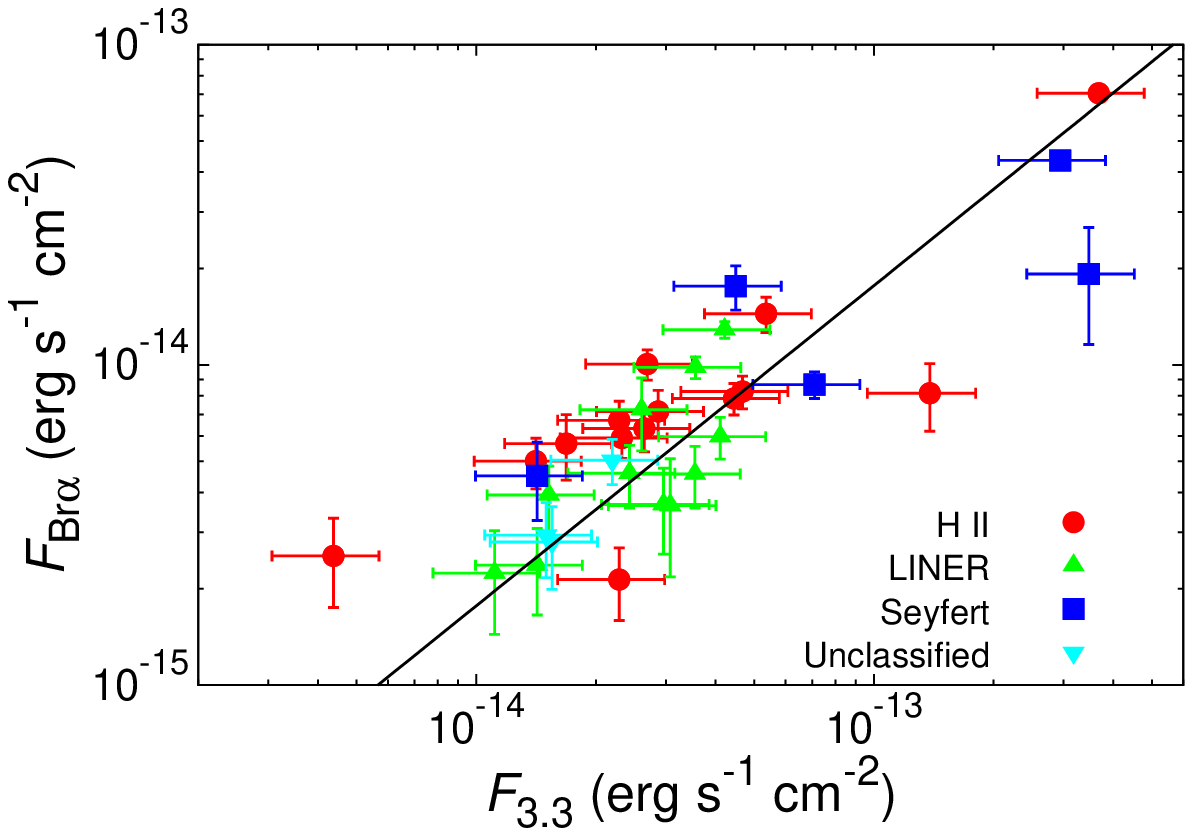}
\caption{
Comparison of
the Br$\alpha$ line with
the 3.3 $\mu$m  PAH emission.
The left panel shows the comparison of the luminosities,
while the right panel presents the comparison of the fluxes.
The symbols and colors represent the optical classifications
of the galaxy.
The solid line shows
the mean $L_{\mathrm{Br}\alpha}/L_{3.3}$
($F_{\mathrm{Br}\alpha}/F_{3.3}$) ratio
of $0.177\pm0.003$.
}
\label{fig:brvspah}
\end{figure*}

The left panel of Figure~\ref{fig:brvspah} shows
the relation between $L_{3.3}$ and $L_{\mathrm{Br}\alpha}$.
We use $L_{3.3}$
listed in Table~\ref{tab:sfcav}.
Dust extinction is corrected for $L_{3.3}$
using the H$\alpha$/H$\beta$ line ratio. 
Extinction at the wavelength of
the 3.3~$\mu$m PAH emission is assumed to
$A_{3.3}=5.32\times10^{-2}A_V$ \citep{Draine2003idg}.
In contrast to the comparison
of $L_{\mathrm{Br}\alpha}$ with $L_\mathrm{IR}$,
$L_{3.3}$ and $L_{\mathrm{Br}\alpha}$ are well correlated with each other,
regardless of the optical classifications of the galaxies.
The correlation coefficient between $L_{3.3}$ and $L_{\mathrm{Br}\alpha}$
is calculated to be 0.659 for a sample size of 33.
The probability of obtaining such a correlation coefficient
by chance is less than $10^{-4}$.
We regard this correlation as statistically significant.
We also examine the comparison of fluxes,
$F_{3.3}$ and $F_{\mathrm{Br}\alpha}$,
in the right panel of Figure~\ref{fig:brvspah}
to investigate a possible correlation introduced by
redshift in the luminosity comparison.
The correlation coefficient between $F_{3.3}$ and $F_{\mathrm{Br}\alpha}$
is 0.917, and the probability of obtaining this value by chance is well below $10^{-4}$.
Thus we conclude that the correlation between the 3.3~$\mu$m PAH
emission and the Br$\alpha$ line is not affected by redshift and is real.

The result of the correlation
between $L_{3.3}$ and $L_{\mathrm{Br}\alpha}$
indicates that they
trace the same excitation sources,
i.e., star formation.
If the broad line region of AGN
contributes to line fluxes,
hydrogen lines would have a FWHM
of a few thousands km~s$^{-1}$
\citep{Osterbrock2006agn}.
However,
the width of the Br$\alpha$ line
is consistent with the spectral resolution
($\Delta v\sim3000$~km~s$^{-1}$)
in all targets
within a fitting uncertainty
of $\la100$~km~s$^{-1}$.
This indicates that
none of the objects show a broad component
of the Br$\alpha$ line
with a FWHM
broader than $\sim1000$~km~s$^{-1}$.
Combining the absence
of the broad component of the Br$\alpha$ line
and the correlation
between $L_{\mathrm{Br}\alpha}$
and $L_{3.3} $,
we assume that the Br$\alpha$ line and the 3.3$\mu$m
PAH emission are
entirely produced by star formation.
The absence of a correlation between 
$L_{\mathrm{Br}\alpha}$ and $L_\mathrm{IR}$
then indicates that $L_\mathrm{IR}$
has a contribution from energy sources other than star formation,
i.e., AGN, in our sample.
Thus we conclude that we are able to
investigate the contribution of starburst as the
dust-enshrouded energy source in ULIRGs
with comparing $L_{\mathrm{Br}\alpha}$ with $L_\mathrm{IR}$.

The 3.3~$\mu$m PAH emission,
which is stronger than the Br$\alpha$ line,
is detected in almost all our targets,
even in the objects in which the Br$\alpha$ line is not observed
(Tab.~\ref{iras}).
Therefore, we calibrate $L_{3.3}$ with $L_{\mathrm{Br}\alpha}$
to quantitatively investigate star formation
in the objects with no Br$\alpha$ line detection.
From the correlation between $L_{3.3}$ with $L_{\mathrm{Br}\alpha}$,
we assume a proportionality between them.
The mean $L_{\mathrm{Br}\alpha}/L_{3.3}$ ratio 
is calculated as
\begin{eqnarray}
L_{\mathrm{Br}\alpha}/L_{3.3}=0.177\pm0.003.
\label{pahtobra}
\end{eqnarray}
This relation quantitatively
associates $L_{3.3}$ 
with the number of ionizing photons from OB stars.
Thus we utilize $L_{3.3}$
as a quantitative indicator of star formation
for galaxies in which the Br$\alpha$ line
is not detected.

When using the 3.3~$\mu$m PAH emission as a proxy of star formation,
\cite{Kim20123pa} pointed out a caveat
that $L_{3.3}/L_\mathrm{IR}$ decreases as
$L_\mathrm{IR}$ increases
within an infrared luminosity range of
$10^{10}L_\odot<L_\mathrm{IR}<10^{13}L_\odot$,
and thus, the power of $L_{3.3}$ as an indicator of
star formation
may be hampered in ULIRGs.
This deficit of $L_{3.3}$ was found
even if the sources likely to be contaminated by AGN activity
are excluded \citep{Yamada2013rp3}.
However, we obtain a 
clear correlation
between $L_{3.3}$
and $L_{\mathrm{Br}\alpha}$.
We attribute this result to the narrowness of the infrared luminosity
range of our sample (most of them lie within
$10^{12}L_\odot<L_\mathrm{IR}<10^{12.5}L_\odot$).
The infrared luminosity of the objects where the Br$\alpha$ line is not detected
is also in this luminosity range, and hence, we assume that
the proportionality of
Equation~(\ref{pahtobra}) is also valid for these objects.

The right panel of Figure~\ref{fig:brvsir} shows the comparison
of $L_{3.3}$ with $L_{\mathrm{IR}}$.
The number of the sample
is increased from 33
in the left panel of the comparison
of $L_{\mathrm{Br}\alpha}$ with $L_\mathrm{IR}$
to 46
with the use of the 3.3 $\mu$m PAH emission.
In this larger sample, we again find that
galaxies show a significant scatter.
The correlation coefficient between $L_\mathrm{IR}$ and $L_{3.3}$
is calculated to be 0.161,
yielding the probability of obtaining such a correlation coefficient
by chance of 29.7\%.
The correlation between $L_\mathrm{IR}$ and $L_{3.3}$
is not significant, as well as that between
$L_\mathrm{IR}$ and $L_{\mathrm{Br}\alpha}$.
We discuss the origin of these scatters in the next section.

\section{DISCUSSION}

In \S\ref{sec:res},
we utilize the luminosities of the Br$\alpha$ line
and the 3.3~$\mu$m PAH emission
as indicators of star formation.
Using these indicators,
we discuss the contribution of starburst
to the total energy from ULIRGs.

\subsection{Starburst Contribution}
\label{test}

The optical classifications of ULIRGs
are mainly based on the optical emission line ratios \citep{Baldwin1981cpe}
and reflect a qualitative difference of the energy sources,
while they have little quantitative information.
We investigate the fractional contribution of starburst
to the total infrared luminosity with the $L_{\mathrm{Br}\alpha}/L_\mathrm{IR}$
and $L_{3.3}/L_\mathrm{IR}$ ratios
and discuss the quantitative difference of the energy sources
among the galaxies with different optical classifications.

\subsubsection{Difference among Galaxies with Different Optical Classifications}
\label{subsec:rdif}

Focusing on the optical classifications of the galaxies,
we find a trend that
the galaxies classified as LINERs or Seyferts
are distributed lower than those
classified as \ion{H}{2} galaxies
in both the panels of Figure~\ref{fig:brvsir}.
This indicates that
LINERs and Seyferts
have lower
$L_{\mathrm{Br}\alpha}/L_{\mathrm{IR}}$ 
and $L_{3.3}/L_\mathrm{IR}$ ratios
than \ion{H}{2} galaxies.
Here we discuss the
significance of these differences.

\begin{figure*}
\plottwo{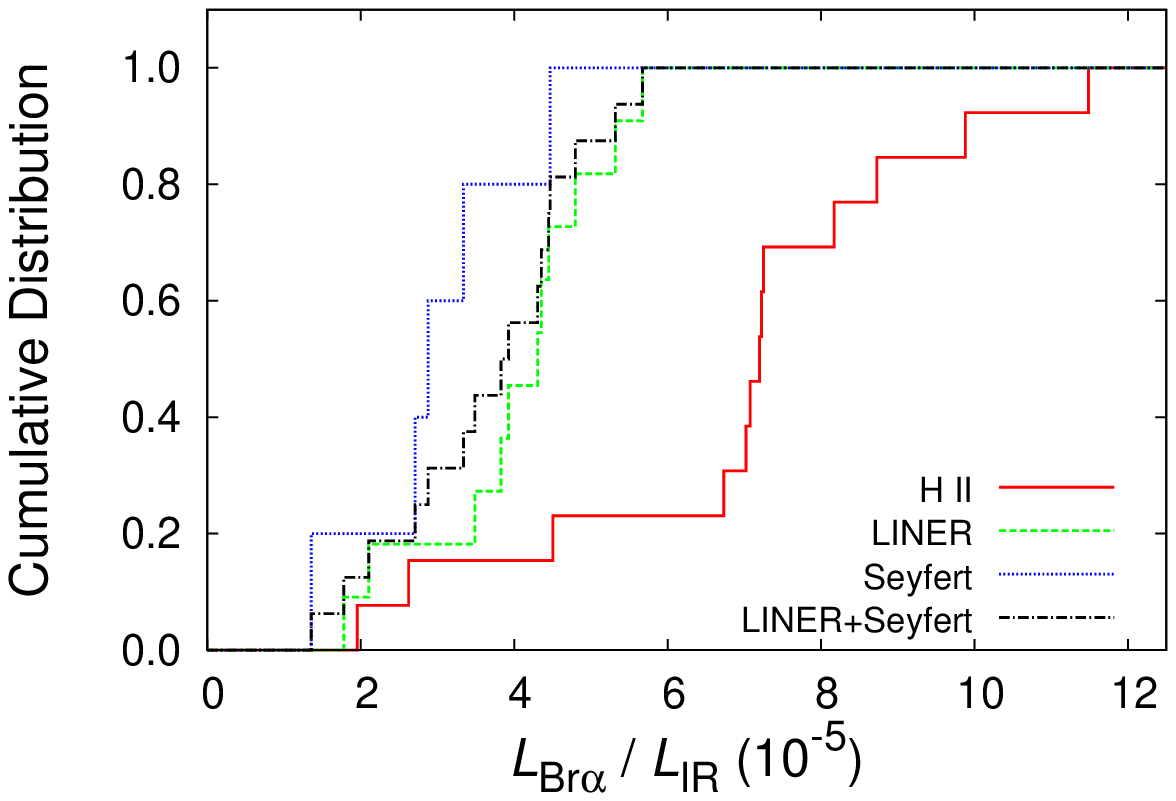}{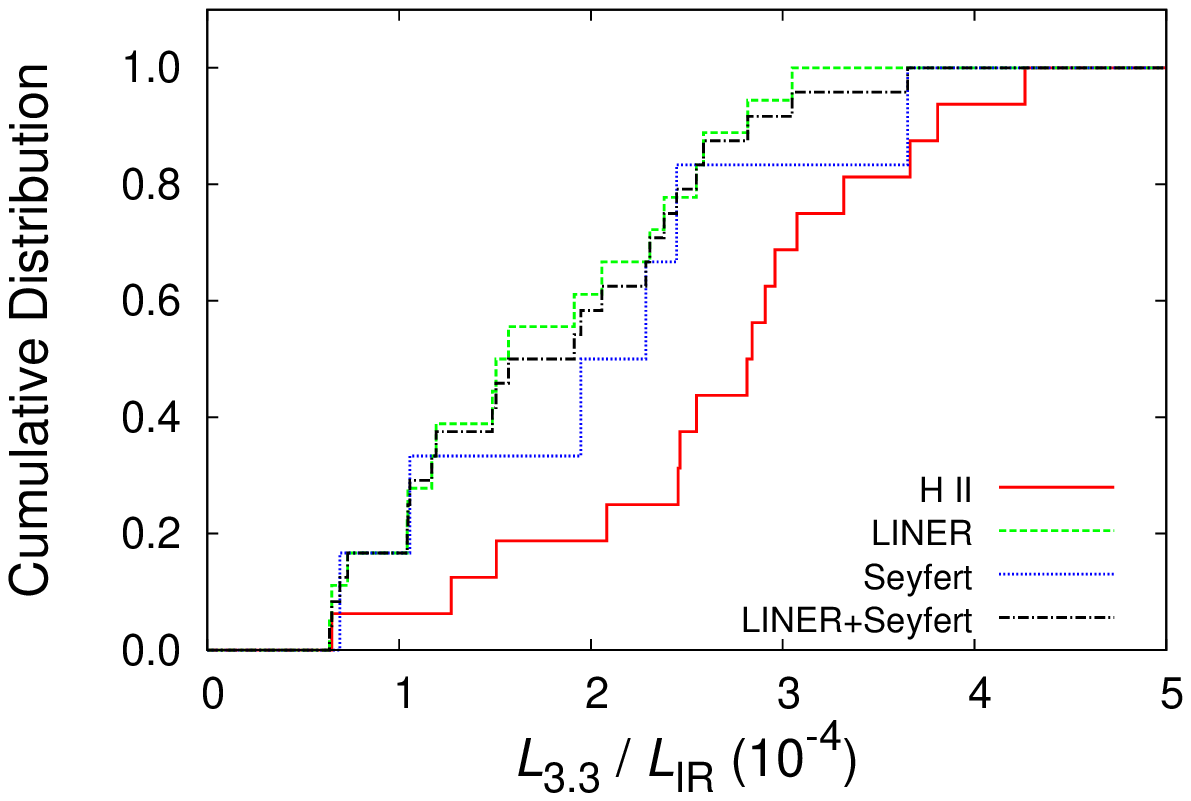}
\caption{
Cumulative distribution for the K--S test of
the $L_{\mathrm{Br}\alpha}/L_\mathrm{IR}$ ratio (left)
and the $L_{3.3}/L_\mathrm{IR}$ ratio (right)
for \ion{H}{2} galaxies (red solid line),
LINERs (green dashed line), Seyferts (blue dotted line),
and the combination of LINERs and Seyferts (black dashed--dotted line).
}
\label{sfcksbr}
\end{figure*}

The left panel of
Figure~\ref{sfcksbr} compares
a distribution of the $L_{\mathrm{Br}\alpha}/L_\mathrm{IR}$ ratio
for \ion{H}{2} galaxies, LINERs, and Seyferts.
The distributions
of LINERs and Seyferts are similar
with each other,
while that of \ion{H}{2} galaxies clearly deviates from the others.
To investigate whether
the difference of the $L_{\mathrm{Br}\alpha}/L_\mathrm{IR}$ ratio
is statistically significant,
we performed the Kolmogorov--Smirnoff (K--S) test
between the galaxies with the different
optical classifications.
The K--S test probabilities that the two sets of samples
originate from the same population
are summarized
in Table~\ref{sfctest}.
The K--S probability between LINERs and Seyferts is
$8.5\times10^{-2}$ and indicates no significant difference
between them.
On the other hand, \ion{H}{2} galaxies show
a probability of less than $5\times10^{-3}$
against LINERs or Seyferts.
If we combine LINERs and Seyferts,
the probability that \ion{H}{2} galaxies and the others
are drawn from the same population is
$1.4\times10^{-4}$.
This difference is statistically significant,
and we conclude that
the $L_{\mathrm{Br}\alpha}/L_{\mathrm{IR}}$ ratio
is statistically different between \ion{H}{2} galaxies and LINERs/Seyferts.

\begin{deluxetable*}{cccccc}			
\tablecaption{Statistical Tests for $L_{\mathrm{Br}\alpha}/L_\mathrm{IR}$ and $L_{3.3}/L_{\mathrm{IR}}$.}
\tablewidth{0pt}
\tablehead{
\colhead{Optical}&\multicolumn{2}{c}{Probability for $L_{\mathrm{Br}\alpha}/L_\mathrm{IR}$}&\colhead{}&\multicolumn{2}{c}{Probability for $L_{3.3}/L_\mathrm{IR}$}\\
\cline{2-3}\cline{5-6}
\colhead{Class}&\colhead{Same Population}&\colhead{Same Mean}&\colhead{}&\colhead{Same Population}&\colhead{Same Mean}\\
\colhead{}&\colhead{(K--S Test)}&\colhead{($t$ Test)}&\colhead{}&\colhead{(K--S Test)}&\colhead{($t$ Test)}
}
\startdata
\ion{H}{2} vs LINER&$6.5\times10^{-4}$&$2.4\times10^{-3}$&&$1.1\times10^{-2}$&$2.8\times10^{-3}$\\
\ion{H}{2} vs Seyfert&$4.1\times10^{-3}$&$5.6\times10^{-3}$&&$6.2\times10^{-2}$&$1.8\times10^{-1}$\\
LINER vs Seyfert&$8.5\times10^{-2}$&$1.2\times10^{-1}$&&$8.1\times10^{-1}$&$4.5\times10^{-1}$\\
\ion{H}{2} vs LINER \& Seyfert&$1.4\times10^{-4}$&$9.0\times10^{-4}$&&$4.0\times10^{-3}$&$3.7\times10^{-3}$
\enddata
\label{sfctest}
\end{deluxetable*}

The right panel of
Figure~\ref{sfcksbr} is the same as the left panel,
but for the $L_{3.3}/L_\mathrm{IR}$ ratio
with the larger sample.
The distribution
of \ion{H}{2} galaxies again deviates from the two other
classifications.
We performed the K--S test
for the $L_{3.3}/L_{\mathrm{IR}}$ ratio
in the same way as the $L_{\mathrm{Br}\alpha}/L_{\mathrm{IR}}$ ratio
and summarize the K--S probabilities 
for the $L_{3.3}/L_{\mathrm{IR}}$ ratio
in Table~\ref{sfctest}.
The K--S test between LINERs and Seyferts 
again indicates no significant difference
between them,
while the probability
between \ion{H}{2} galaxies
and the combination of LINERs and Seyferts
is $4.0\times10^{-3}$.
This means that
the distribution of the $L_{3.3}/L_\mathrm{IR}$
is statistically different between \ion{H}{2} galaxies
and LINERs/Seyferts.
Thus the result we obtained with
the $L_{\mathrm{Br}\alpha}/L_{\mathrm{IR}}$ ratio
is reproduced using $L_{3.3}$ with the larger sample.

The mean $L_{\mathrm{Br}\alpha}/L_\mathrm{IR}$ ratio
in each classification is calculated and shown in Table~\ref{sfcmean}.
The combination of LINERs and Seyferts
yields the mean $L_{\mathrm{Br}\alpha}/L_\mathrm{IR}$ ratio
of $3.7\times10^{-5}$, which is about half of that
in \ion{H}{2} galaxies ($6.9\times10^{-5}$).
To investigate whether the mean
$L_{\mathrm{Br}\alpha}/L_{\mathrm{IR}}$
ratio in LINERs or Seyferts
is statistically lower than that in \ion{H}{2} galaxies,
we performed the Student's $t$ test.
The $t$-test probabilities that two sets of samples
originate from populations with the same mean 
$L_{\mathrm{Br}\alpha}/L_{\mathrm{IR}}$ ratio
are summarized
in Table~\ref{sfctest}.
The test between LINERs and Seyferts
indicates that there is no significant difference between them,
whereas \ion{H}{2} galaxies show
a low probability
against the others.
The probability between \ion{H}{2} galaxies and the
combination of LINERs and Seyferts is
$9.0\times10^{-4}$, which is statistically significant.
Thus, we conclude that the mean
$L_{\mathrm{Br}\alpha}/L_{\mathrm{IR}}$
ratio in LINERs and Seyferts
is significantly lower, about half
of that in \ion{H}{2} galaxies.

\begin{deluxetable*}{cccccccc}				
\tablecaption{Mean and Deviation of $L_{\mathrm{Br}\alpha}/L_{\mathrm{IR}}$ and $L_{3.3}/L_{\mathrm{IR}}$.}
\tablewidth{0pt}
\tablehead{
\colhead{Optical}&\multicolumn{3}{c}{$L_{\mathrm{Br}\alpha}/L_{\mathrm{IR}}$}&\colhead{}&\multicolumn{3}{c}{$L_{3.3}/L_{\mathrm{IR}}$}\\\cline{2-4}\cline{6-8}
\colhead{Class}&\colhead{Number}&\colhead{Mean}&\colhead{Deviation\tablenotemark{a}}&\colhead{}&\colhead{Number}&\colhead{Mean}&\colhead{Deviation\tablenotemark{a}}\\
\colhead{}&\colhead{of Objects}&\colhead{($10^{-5}$)}&\colhead{($10^{-5}$)}&\colhead{}&\colhead{of Objects}&\colhead{($10^{-4}$)}&\colhead{($10^{-4}$)}
}
\startdata
\ion{H}{2}&13&6.9&2.6&&16&2.7&0.9\\
LINER&11&4.0&1.2&&18&1.7&0.8\\
Seyfert&5&3.0&1.1&&6&2.0&1.1\\
LINER \& Seyfert&16&3.7&1.2&&24&1.8&0.8
\enddata
\tablenotetext{a}{1-$\sigma$ standard deviation.}
\label{sfcmean}
\end{deluxetable*}

In the same way as
the $L_{\mathrm{Br}\alpha}/L_{\mathrm{IR}}$ ratio,
the difference of the mean $L_{3.3}/L_\mathrm{IR}$ ratio
(Table~\ref{sfcmean}) between the galaxies
with different optical classifications is examined
with the Student's $t$ test.
The  result is summarized in Table~\ref{sfctest}.
From the $t$ test probabilities for the $L_{3.3}/L_\mathrm{IR}$ ratio, 
we conclude that the mean $L_{3.3}/L_\mathrm{IR}$ ratio
in LINERs and Seyferts
is significantly lower than that in \ion{H}{2} galaxies.
This is consistent with the result obtained from
the $L_{\mathrm{Br}\alpha}/L_{\mathrm{IR}}$ ratio.
Thus, with the larger sample,
the difference
among the galaxies
with different optical classifications is further confirmed.

We here discuss the possible effect
of dust extinction to the difference
of the $L_{\mathrm{Br}\alpha}/L_{\mathrm{IR}}$
and $L_{3.3}/L_{\mathrm{IR}}$ ratios.
To explain the difference of the mean
$L_{\mathrm{Br}\alpha}/L_{\mathrm{IR}}$
or $L_{3.3}/L_{\mathrm{IR}}$ ratio
among the galaxies
with different optical classifications
with dust extinction,
extinction should be much higher
in LINERs and Seyferts than in \ion{H}{2} galaxies.
However, \cite{Veilleux2009sqa} reported
that the optical depth of the 9.7~$\mu$m silicate
absorption was generally smaller in Seyferts than
in \ion{H}{2} galaxies based on the \textit{Spitzer} results.
This indicates that the dust extinction in \ion{H}{2}
galaxies is generally higher than that in Seyferts
and is opposite to the above scenario.
Therefore,
we conclude that
the difference of the mean
$L_{\mathrm{Br}\alpha}/L_{\mathrm{IR}}$
and $L_{3.3}/L_{\mathrm{IR}}$ ratios
among
the galaxies
with different optical classifications
cannot be explained by
the effect of heavy dust extinction.

In summary,
we conclude that the mean $L_{\mathrm{Br}\alpha}/L_\mathrm{IR}$
and $L_{3.3}/L_\mathrm{IR}$ ratios are
significantly lower in LINERs and Seyferts
than in \ion{H}{2} galaxies.
This difference is not attributable to 
the effect of dust extinction.

\subsubsection{Fractional Contribution of Starburst
as Energy Sources}
\label{subsec:fra}

We propose that
the $L_{\mathrm{Br}\alpha}/L_\mathrm{IR}$
and $L_{3.3}/L_\mathrm{IR}$ ratios
reflect the fractional contribution of starburst to
the total infrared luminosity,
and the relative difference of the ratios
among the galaxies with the different optical classifications
shows the difference of the energy sources of them.

We here assume that \ion{H}{2} galaxies are completely
energized by starburst,
and the $L_{\mathrm{Br}\alpha}/L_\mathrm{IR}$ ratio
of $6.9\times10^{-5}$,
which is the mean ratio in \ion{H}{2} galaxies,
corresponds to the starburst contribution of 100\%.
On the basis of this assumption,
we estimate the starburst contribution
as $R_\mathrm{SB}^\mathit{AKARI}=
(L_{\mathrm{Br}\alpha}/L_\mathrm{IR})
/(L_{\mathrm{Br}\alpha}/L_\mathrm{IR})_0$,
where $(L_{\mathrm{Br}\alpha}/L_\mathrm{IR})_0=6.9\times10^{-5}$.

\begin{figure}
\plotone{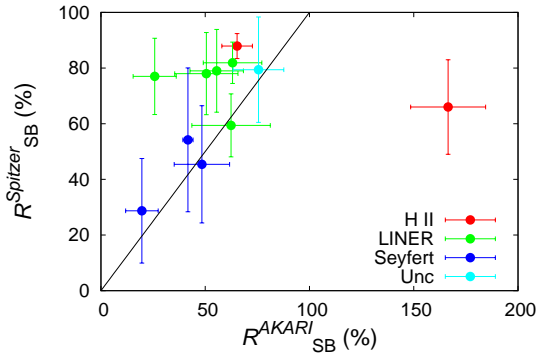}
\caption{
Comparison of
the starburst contribution to the bolometric luminosity of the galaxy
measured by the \textit{Spitzer} result
($R_\mathrm{SB}^\mathit{Spitzer}$)
with that
estimated from
our \textit{AKARI} result
($R_\mathrm{SB}^\mathit{AKARI}$; see \S\ref{subsec:rdif}).
$R_\mathrm{SB}^\mathit{Spitzer}$ is calculated from
Column~8 of Table~12 in \cite{Veilleux2009sqa}.
The error bar of $R_\mathrm{SB}^\mathit{Spitzer}$ represents
the standard deviation of the six methods
of Table~12 in \cite{Veilleux2009sqa}.
The symbols and colors represent the optical classifications
of the galaxy.
The black solid line indicates $R_\mathrm{SB}^\mathit{Spitzer}
=R_\mathrm{SB}^\mathit{AKARI}$.
}
\label{spitzeragn}
\end{figure}

Even the \textit{AKARI} observations, however, could be affected by 
dust extinction.
To verify this effect, we compare our result with the 
mid-infrared spectroscopic results obtained by \textit{Spitzer}
IRS spectroscopy, which is expected 
to be less affected by dust extinction.
\cite{Veilleux2009sqa}
summarized the six independent
diagnostics to estimate
the starburst/AGN contribution in ULIRGs
based on \textit{Spitzer} observations.
Among our targets,
11 galaxies are also reported in
\cite{Veilleux2009sqa},
and the starburst contribution to the bolometric luminosity
($R_\mathrm{SB}^\mathit{Spitzer}$)
in these galaxies is calculated.
To see the consistency of our work and
the previous \textit{Spitzer} works,
we compare $R_\mathrm{SB}^\mathit{AKARI}$
with $R_\mathrm{SB}^\mathit{Spitzer}$
in Figure~\ref{spitzeragn}.

Figure~\ref{spitzeragn} shows general agreement between 
$R_\mathrm{SB}^\mathit{AKARI}$
and $R_\mathrm{SB}^\mathit{Spitzer}$,
but we note
$R_\mathrm{SB}^\mathit{Spitzer}$ might be
slightly higher than
$R_\mathrm{SB}^\mathit{AKARI}$
especially in LINERs.
To investigate the difference
of $R_\mathrm{SB}^\mathit{Spitzer}$ and
$R_\mathrm{SB}^\mathit{AKARI}$,
we estimate the error-weighted mean
$R_\mathrm{SB}^\mathit{Spitzer}/
R_\mathrm{SB}^\mathit{AKARI}$ ratio.
The ratio of the total sample
is  $0.84\pm0.15$, while
it is  $0.69\pm0.44$,
$1.13\pm0.15$,
and $1.28\pm0.16$
in \ion{H}{2} galaxies, Seyferts, and LINERs, respectively.
The biggest deviation of the ratio from unity
is seen in LINERs with the level of  
about 30\%, 
but the significance level is not high ($1.8\sigma$).
Due to the limited number of samples, 
we cannot deny the possibility that the two
results might be different, but
the current sample does not show the clear difference between the
starburst contributions estimated by the two methods.
We hence conclude that
our result is consistent with the \textit{Spitzer} result,
and so $R_\mathrm{SB}^\mathit{AKARI}$
can be used as a quantitative indicator
of the starburst contribution
to the total infrared luminosity.
We discuss below the energy sources of the galaxies
using the $L_{\mathrm{Br}\alpha}/L_\mathrm{IR}$ ratio.

We should
note that the mean $R_\mathrm{SB}^\mathit{Spitzer}/R_\mathrm{SB}^\mathit{AKARI}$
ratio in the entire sample is
affected by one outlying \ion{H}{2} galaxy, IRAS~17068$+$4027.
If we calculate the mean
$R_\mathrm{SB}^\mathit{Spitzer}/R_\mathrm{SB}^\mathit{AKARI}$
ratio without IRAS~17068$+$4027,
the ratio becomes 1.30$\pm$0.08 
(c.f., $0.84\pm0.15$ with IRAS~17068$+$4027), 
and so our \textit{AKARI} observation
could underestimate the starburst contribution
about 30\% relative to the \textit{Spitzer} observation.
We have searched possible cause of the high
$R_\mathrm{SB}^\mathit{AKARI}$
value relative to $R_\mathrm{SB}^\mathit{Spitzer}$
in IRAS~17068$+$4027, 
but have not successfully identified
the cause.
For example, we cannot attribute this to the hidden AGN activity,
since Figure~\ref{spitzeragn} shows that
the $R_\mathrm{SB}^\mathit{Spitzer}/
R_\mathrm{SB}^\mathit{AKARI}$ ratio
of Seyferts, which are dominated by AGN,
is not deviated from unity.
Moreover, we cannot find any clear evidence
of the presence of AGN in this galaxy
from our observation (e.g., absence of the broad line)
and other literature \citep[e.g.,][]{Veilleux1997n,Kim1998i1j}.
We hence include this galaxy in the estimate of
$R_\mathrm{SB}^\mathit{Spitzer}/R_\mathrm{SB}^\mathit{AKARI}$
since no clear reason
to ignore this galaxy has been found.

\subsubsection{Energy Sources of ULIRGs}

Starburst is thought to be the dominant
energy source in \ion{H}{2} galaxies, while
AGN are considered to dominate in Seyferts.
The result that the mean
$L_{\mathrm{Br}\alpha}/L_{\mathrm{IR}}$ ratio
is lower in
Seyferts
than in
\ion{H}{2} galaxies
confirms
this difference of energy sources.
The mean $L_{\mathrm{Br}\alpha}/L_\mathrm{IR}$ ratio in
Seyferts is about 43\% of that in \ion{H}{2} galaxies.
This means that
the fractional contribution of starburst 
to the total infrared luminosity in Seyferts is $\sim43\%$.
Given the result that LINERs also
have a lower ($\sim58\%$)
$L_{\mathrm{Br}\alpha}/L_{\mathrm{IR}}$ ratio
than \ion{H}{2} galaxies,
energy sources other than starburst
are needed in LINERs.
The $L_{\mathrm{Br}\alpha}/L_{\mathrm{IR}}$ ratio
in LINERs is similar to that in Seyferts,
and thus, we propose that
AGN are needed as
energy generation mechanisms in LINERs,
as well as Seyferts.
The $L_{3.3}/L_{\mathrm{IR}}$ ratio
in LINERs is also similar to that in Seyferts
and lower than that in \ion{H}{2} galaxies.
Thus the idea that
AGN are needed in LINERs, as well as Seyferts,
is also supported in the larger sample.

The combination of LINERs and Seyferts
yields the mean $L_{\mathrm{Br}\alpha}/L_\mathrm{IR}$ ratio
of $3.7\times10^{-5}$, which indicates that
the fractional contribution of starburst
to the total infrared luminosity in these galaxies is $\sim50\%$.
With the larger sample,
using the $L_{3.3}/L_\mathrm{IR}$ ratio,
the contribution of starburst to the total infrared luminosity
in the combination of LINERs and Seyferts
is estimated as $\sim67\%$.
From this result, it is inferred that
AGN produce a significant fraction ($\sim33\%$)
of the total infrared luminosity in LINERs and Seyferts,
causing the separation of
the ratios
between \ion{H}{2} galaxies and LINERs/Seyferts.
As a whole, our result indicates that
AGN account for about one-third
of the total infrared luminosity
of LINERs and Seyferts, and 
starburst explains the remaining infrared luminosity of ULIRGs.
The mean $L_{3.3}/L_\mathrm{IR}$ ratio
in the entire sample is calculated as $2.1\times10^{-4}$,
which yields the contribution of starburst
to the total infrared luminosity of $\sim80\%$.
Consequently, the AGN contribution to the total infrared luminosity
in the entire sample is indicated as $\sim20\%$.
Thus, we conclude that starburst is the dominant power
source for the extreme infrared luminosity of ULIRGs.

In addition to \cite{Veilleux2009sqa},
numerous works have been done
using the mid-infrared spectroscopy with
the \textit{Spitzer} satellite
to reveal the dust-enshrouded energy sources of
ULIRGs.
\cite{Armus2007oui} investigated
the mid-infrared spectra of 10 ULIRGs
and found evidence for AGNs in galaxies with optical
Seyfert or LINER classifications,
while they did not find evidence
for buried AGNs in ULIRGs classified
optically as \ion{H}{2} galaxies.
\cite{Nardini2009eag} analyzed 5--8~$\mu$m
spectra of 71 ULIRGs and derived
AGN/starburst contribution to
the overall energy output of each source
using the spectral decomposition technique.
They found that the main fraction
of ULIRG luminosity arose from starburst,
while the AGN contribution
was non-negligible ($\sim$23$\%$).
From radio observations of 7 ULIRGs,
\cite{Prouton2004mse}
concluded that the AGN contribution
was at most 50\%.
Our conclusion that
starburst is the dominant power
source for ULIRGs is consistent with
these results of the works at the longer wavelength.
In particular,
our estimation of the AGN contribution
to the total infrared luminosity in our entire sample ($\sim20\%$)
shows a good agreement with the
result obtained by \cite{Nardini2009eag}.

As for the less luminous population,
luminous infrared galaxies
(LIRGs; $L_\mathrm{IR}=10^{11}$--$10^{12}L_\odot$),
\cite{Alonso-Herrero2012lli} decomposed
the \textit{Spitzer} spectra of 53 LIRGs into
AGN and starburst components.
They found that the AGN contribution
was only 5$\%$ on average in their sample
and the bulk of the infrared luminosity of these LIRGs
was due to the starburst activity.
Our estimation of the AGN contribution
in ULIRGs is larger than this result of the
LIRGs observation.
As discussed in \cite{Alonso-Herrero2012lli},
this supports the idea that
the AGN contribution to the total infrared luminosity
increases with the total infrared luminosity.
\cite{Shipley2013ssi} investigated
mid-infrared PAH emissions in 65 LIRGs
and estimated the AGN contribution to
the total infrared luminosity in their targets.
They divided their sample into
a subsample of galaxies with \textit{Spitzer}
3.6--8.0~$\mu$m colors
indicative of warm dust heated
by AGN (IRAGN; 14 galaxies)
and those galaxies whose colors indicated
starburst processes (non-IRAGN; 65 galaxies).
They found that for most
IRAGN starburst accounted for 10$\%$--50$\%$
of the total IR luminosity,
while non-IRAGN were mostly dominated by starburst.
Their estimation of the starburst contribution
in IRAGN is quite low.
It is even lower than our estimation of the starburst contribution
in LINERs and Seyferts ($\sim67\%$).
We attribute the low starburst contribution of IRAGN
relative to our estimation
to the difference of the classification methods of galaxies.
We propose that the \textit{Spitzer} classification method of IRAGN
separated AGN dominated galaxies better than the optical classification method,
which should suffer from the effect of dust extinction,
and caused the low starburst contribution in IRAGN sample.

The longer wavelength results
rely on a little complicated
indicators such as fine structure lines
or spectral decomposition techniques
which assume empirical starburst and AGN templates.
On the other hand,
our method is based on the direct indicator
of the ionizing photons, the Br$\alpha$ line,
and so is able to estimate the contribution
of starburst in a robust way.
We conclude that 
the contribution of starburst to the total infrared luminosity
is different among the galaxies with different optical classifications.
The starburst contribution is estimated as $\sim67\%$ in LINERs and Seyferts.
Our result is consistent with the previous
works at the longer wavelength.

\subsection{Deficit of Ionizing Photons}
\label{subsec:def}

Here we revisit the assumption
of 100\% contribution of starburst in \ion{H}{2} galaxies with
converting the observed luminosities to
the number of ionizing photons.
The Br$\alpha$ line luminosity can be converted
into the number of ionizing photons,
$Q_{\mathrm{Br}\alpha}$, on the assumption of the
case B with $T=10000$~K and low-density limit
of the model by \cite{Osterbrock2006agn};
\begin{eqnarray}
Q_{\mathrm{Br}\alpha}\ (\mathrm{s}^{-1})
=&2.54\times10^{13}L_{\mathrm{Br}\alpha}\ (\mathrm{erg\ s}^{-1}).
\label{ltoq}
\end{eqnarray}
The number of ionizing photons from OB stars
is theoretically related to SFR
on the assumption of the initial mass function
\citep[e.g.,][]{Kennicutt2012sfi}.
On the other hand,
the total infrared luminosity can also be converted to SFR
on the assumption of the initial mass function if it is solely generated by star formation
\citep{Kennicutt2012sfi}.
We assume that this is the case in \ion{H}{2} galaxies.
Thus,
we can estimate the number of ionizing
photons ($Q_\mathrm{IR}$) expected from SFR
required to explain the total infrared luminosity.
Adopting
the calibration provided by \cite{Murphy2011cef},
we convert the total infrared luminosity
into the number of ionizing photons as
\begin{eqnarray}
Q_\mathrm{IR}\ (\mathrm{s}^{-1})=5.33\times10^{9}L_\mathrm{IR}\ (\mathrm{erg\ s}^{-1}).
\label{sfrir}
\end{eqnarray}
The $L_{\mathrm{Br}\alpha}/L_\mathrm{IR}$ ratio of
$2.1\times10^{-4}$ is required to obtain
$Q_{\mathrm{Br}\alpha}/Q_\mathrm{IR}=100\%$.

Using Equations~(\ref{ltoq}) and (\ref{sfrir}),
we convert the $L_{\mathrm{Br}\alpha}/L_\mathrm{IR}$
ratio into the $Q_{\mathrm{Br}\alpha}/Q_\mathrm{IR}$ ratio.
We find that the mean $L_{\mathrm{Br}\alpha}/L_\mathrm{IR}$
ratio in \ion{H}{2} galaxies of $6.9\times10^{-5}$
yields the $Q_{\mathrm{Br}\alpha}/Q_\mathrm{IR}$ ratio
of only $\sim33\%$.
This indicates that starburst
explains merely less than one-third of
the total infrared luminosity,
even in \ion{H}{2} galaxies.
This is inconsistent with the
assumption
that \ion{H}{2} galaxies are dominated by starburst.
Similar results that the number of ionizing
photons derived from
the near-infrared hydrogen recombination lines
is low relative to that
expected from the total infrared luminosity
in ULIRGs have been reported by \cite{Goldader1995sli} using
the Br$\gamma$ line and \cite{Valdes2005nsl}
using the Pa$\alpha$ and Br$\gamma$ lines.
Our result indicates that the Br$\alpha$ line
also suffers the same kind of deficit.

Comparing our result with the \textit{Spitzer} result (Figure~\ref{spitzeragn}),
we show that $R_\mathrm{SB}^\mathit{AKARI}$,
which is based on the assumption of 100\% contribution of starburst in \ion{H}{2} galaxies,
is consistent with 
the fractional contribution of starburst
estimated from the longer wavelength result (\S\ref{subsec:fra}).
This indicates that 
the apparently low $Q_{\mathrm{Br}\alpha}/Q_\mathrm{IR}$ ratio
is caused by underestimation of the number of ionizing photons with the Br$\alpha$ line by a factor of $\sim3$.

As we see in \S\ref{subsec:hde} with the entire sample,
the Br$\alpha$ line flux could be underestimated by a factor of two
due to dust extinction.
To closely investigate this effect,
we calculate the mean
$Q_{\mathrm{Br}\alpha}/Q_\mathrm{IR}$ ratio
in \ion{H}{2} galaxies
with correcting $L_{\mathrm{Br}\alpha}$
using $A_V^{9.7}$.
We find that
the mean $Q_{\mathrm{Br}\alpha}/Q_\mathrm{IR}$ ratio
increases to $55.5\%$
but is still lower than 100\%.
The standard error 
is estimated as 7.5\% (1$\sigma$),
and so the significance of the deviation from 100\%
is estimated as $\sim6\sigma$ even after the dust extinction correction
with $A_V^{9.7}$.
Thus we conclude that
the number of ionizing photons traced by
the Br$\alpha$ line is deficient relative to
that expected from the total infrared luminosity
even taking heavy dust extinction into consideration.

In addition to heavy dust extinction,
we propose that
dust within the starburst ionized
regions absorbs
a significant fraction of ionizing photons from OB stars
and causes
the underestimation of
the number of ionizing photons with the Br$\alpha$ line.
If only a fraction $f\%$ of photons with energy $>13.6$~eV
ionizes the gas while the remaining $(100-f)\%$ is absorbed by dust,
the hydrogen lines underestimate the number of ionizing photons to be $f$\% of the intrinsic value.
\cite{Hirashita2003sfr} have estimated an average value of
$(100-f)\sim50\%$, with some objects reaching $(100-f)\sim80\%$,
based on observations of IUE-selected star-forming galaxies.
To explain the discrepancy between
$Q_{\mathrm{Br}\alpha}$
and $Q_\mathrm{IR}$
in \ion{H}{2} galaxies
after taking the uncertainty of dust extinction
into consideration,
the fraction of ionizing photons absorbed by dust
is required to be $\sim45\%$,
which is well within the range of the possible values
indicated by \cite{Hirashita2003sfr}.
We suggest that
the absorption of ionizing photons
within \ion{H}{2} regions
is required in addition to heavy dust extinction to
the Br$\alpha$ line to explain
the underestimation of $Q_{\mathrm{Br}\alpha}$
relative to $Q_\mathrm{IR}$ in ULIRGs.

\section{SUMMARY}
We conducted systematic observations
of the hydrogen Br$\alpha$ line
with the \textit{AKARI} IRC 2.5--5.0~$\mu$m spectroscopy
in 50 nearby ($z<0.3$) ULIRGs
to estimate
the strength of starburst unbiased by dust extinction.
We detected the Br$\alpha$ line in 33 ULIRGs.
Comparing the Br$\alpha$ line with
the 3.3~$\mu$m PAH emission
and the total infrared luminosity,
we investigate the fractional contribution of starburst
to the total infrared luminosity in ULIRGs.
The main results are as follows:

\begin{enumerate}
\item
The effect of heavy dust extinction to the Br$\alpha$ line is investigated with
the Br$\beta$/Br$\alpha$ line ratio in one source
and the optical depth of 9.7~$\mu$m silicate absorption taken
from \textit{Spitzer} results in the entire sample.
We conclude that the
intrinsic Br$\alpha$ line flux
is determined within the uncertainty of
a factor of two underestimation even
if we take the effect of heavy dust extinction into consideration.
\item
From the comparison of the Br$\alpha$ line luminosity ($L_{\mathrm{Br}\alpha}$)
with the 3.3~$\mu$m PAH emission luminosity ($L_{3.3}$),
we find a good correlation between $L_{3.3}$ and $L_{\mathrm{Br}\alpha}$.
This indicates that $L_{3.3}$ and $L_{\mathrm{Br}\alpha}$
trace the same excitation sources, i.e., star formation.
On the other hand,
the total infrared luminosity ($L_\mathrm{IR}$) and $L_{\mathrm{Br}\alpha}$
show no clear correlation with each other.
Thus a contribution from energy sources other than star formation,
i.e., AGN, to $L_\mathrm{IR}$ is indicated in our sample.
\item
To investigate star formation in fainter objects,
we derive following relation
on the assumption of a proportionality between $L_{3.3}$
and $L_{\mathrm{Br}\alpha}$;
$L_{\mathrm{Br}\alpha}/L_{3.3}=0.177\pm0.003$.
Using $L_{3.3}$ as an indicator of star formation,
we reconfirm that
$L_{3.3}$ and $L_\mathrm{IR}$ show no clear
correlation with each other,
as well as $L_{\mathrm{Br}\alpha}$ and $L_\mathrm{IR}$,
in a larger sample of 46 galaxies
in which objects with no Br$\alpha$ line detection
are included.
\item
The mean $L_{\mathrm{Br}\alpha}/L_\mathrm{IR}$ ratio
is significantly lower in galaxies optically classified as LINERs and Seyferts
than in \ion{H}{2} galaxies.
This difference is also confirmed
with the $L_{3.3}/L_\mathrm{IR}$ ratio
in the larger sample.
We propose that the difference reflects
the contribution of starburst to the total infrared luminosity in ULIRGs.
Assuming that \ion{H}{2} galaxies are 100\% energized by starburst,
we estimate that the contribution of starburst to the total infrared
luminosity in LINERs and Seyferts is $\sim67\%$,
and active galactic nuclei contribute to
the remaining $\sim33\%$.
\item
We find that
the number of ionizing photons
derived from the Br$\alpha$ line ($Q_{\mathrm{Br}\alpha}$)
are significantly smaller than
that expected from star formation rate required
to explain the total infrared luminosity ($Q_\mathrm{IR}$).
The mean $L_{\mathrm{Br}\alpha}/L_\mathrm{IR}$
ratio in \ion{H}{2} galaxies
yields the $Q_{\mathrm{Br}\alpha}/Q_\mathrm{IR}$ ratio
of only $55.5\pm7.5\%$
even after taking heavy dust extinction into consideration.
We attribute this apparently
low ratio to the underestimation
of the number of ionizing photons with the Br$\alpha$ line.
We conclude that
the number of ionizing photons traced by
the Br$\alpha$ line is deficient relative to
that expected from the total infrared luminosity
even taking the effect of
heavy dust extinction into consideration.
As an additional cause of the deficit,
we propose that dust within \ion{H}{2} regions
absorbs a significant fraction ($\sim45\%$)
of ionizing photons.
\end{enumerate}

\acknowledgments

This work is based on the observations made with
\textit{AKARI}, a JAXA project, with the participation of ESA.
This research also made use of
the NASA/IPAC Extragalactic Database, which is operated
by the Jet Propulsion Laboratory, California Institute of
Technology, under contract with the National Aeronautics and
Space Administration.
We thank the anonymous referee
for his/her useful comments which significantly improved this paper.
This work is supported by JSPS KAKENHI Grant Number 26247030.
K.Y.~is supported through the Leading Graduates
Schools Program, gAdvanced Leading Graduate Course for Photon Science,h by the Ministry of
Education, Culture, Sports, Science and Technology of Japan.

\textit{Facility:} \facility{Akari (IRC)}

\end{document}